\documentclass{aastex}
\usepackage{spr-astr-addons}
\usepackage{url}
\usepackage{subfigure}

\RequirePackage{color}

\newcommand{\emaila}{antonellalucia.iannella@unisannio.it}

\begin{document}

\title{A note on the interpretation of the statistical analysis of the $M_{\bullet}-M_{G}\sigma^2$ scaling relation}

\author{A. L. Iannella\altaffilmark{1}}
\email{antonellalucia.iannella@unisannio.it}
\email{\emaila}
\and
\author{L. Greco\altaffilmark{2}}
\email{l.greco@unifortunato.eu}
\and
\author{A. Feoli\altaffilmark{1}}
\email{feoli@unisannio.it}

\altaffiltext{1}{Department of Engineering, University of Sannio, Piazza Roma 21, 82100 Benevento, Italy}
\altaffiltext{2}{Giustino Fortunato University, Viale Raffaele Delcogliano 12, 82100 Benevento, Italy\\
Corresponding author: A. L. Iannella - antonellalucia.iannella@unisannio.it}

\begin{abstract}
In the context of scaling relations between Supermassive Black Holes and host-galaxy properties, we aim to enhance the comparison between $M_{\bullet} - M_{G}\sigma^2$ and $M_{\bullet} - \sigma$ relations from a statistical point of view. First, it is suggested to take into account the predictive accuracy of the scaling relation, in addition to the classical measures of goodness of fit.  Here, prediction accuracy is fairly evaluated according to a leave-one-out cross-validation strategy. Then, we spread more light on the analysis of residuals from the fitted scaling relation, in order to provide more useful information on the role played by the different variables in their correlation with the black hole mass.
The findings from six samples are discussed.
\end{abstract}


\maketitle

\section{Introduction}

It is well ascertained that at the center of local galaxies there is a supermassive black hole (SMBH, $M_{\bullet}>10^6 M_{\odot}$) \citep{fer2005,kor1995,ric1998}. The SMBH growth and the bulge formation regulate each other \citep{ho2004}. The association between the mass of a SMBH with the properties of the hosting galaxy has been extensively studied in the literature. In particular, the interest focused on the study of the correlation with characteristics such as the bulge luminosity or mass, the velocity dispersion, the effective radius, the Sérsic index, the kinetic energy, etc. \citep{all2007,ber2013,burkert10,feo2005,feo2007,fer2000,geb2000,geb2003,gra2001,
gra2005,gra2007,gul2009,har2004,kor1995,lao2001,lau2007,mag1998,mar2001,mar2003,
mer2001,ric1998,sei2008,snyder11,soker10,tre2002,van1999,wan2002}.
Among all the proposed relations, we studied carefully the performances of the correlation with kinetic energy, as first found by Feoli and Mele in 2005 \citep{feo2005}. The reason is that this relation exhibits solid theoretical foundations and can play the role of an HR diagram for galaxies \citep{feo2009,man2012}. Furthermore, it has already shown to predict the masses of SMBH better than other relations in some particular cases \citep{ben2013}.
In this paper, the objective is to analyze the predictive power of the scaling relation between the mass of a SMBH and kinetic energy and to enhance the statistical analyses in \citet{ian2020}. Then, we will follow  two main paths of investigation:
\begin{itemize}
  \item[1.] we consider the predictive reliability of the scaling relations $M_{\bullet}- M_{G}\sigma^2$ and $M_{\bullet}- \sigma$, that is fairly measured by \textit{leave-one-out cross-validation}, as described in Sect. 3, in addition to the classical goodness of fit measures;

     \vspace{5pt}
  \item[2.] we discuss more in depth the analysis of residuals from the fitted scaling relations. In particular, in Sect. 4 we give an appropriate understanding of the comparison among correlations developed in \citet{bar2017}, \citet{ber2007}, \citet{hop2007a,hop2007b} and \citet{sha2016,sha2019}, aimed to measure the relative importance of the variables that may be included in the scaling relation.

\end{itemize}
The analysis refers to six samples, that are listed in Sect. 2: five samples have been already considered in \citet{ian2020} and one comes from \citet{kor2013}.
The results about goodness of fit and prediction accuracy of the
scaling relations are presented and discussed in Sect. 5. The residual analyses for each sample  are considered in Sect. 6, that provide important information on the role played by the different variables in the correlations under investigation. Some concluding remarks follow in Sect. 7.

\section{Samples}
\label{sample}
The analyses concern the five samples from \citet{ian2020} and one more sample collected in \citet{kor2013}. The results regarding the first five samples are expected to improve over the findings from Feoli and Mele (2005) and \citet{ian2020}. In contrasts, the last sample has not been studied yet, because we planned to include it in an upcoming work related to the role of pseudobulges. Moreover, it is not our intention to modify or comment the results by \citet{kor2013} (see also \citet{sag2016}), but we only aim to offer more insights on some aspects of the statistical analysis. All the six samples have been made available at \textit{ http://people.ding.unisannio.it/feoli/IF2020.zip}.

Here, a brief description of the main characteristics of the six samples follows:
\begin{itemize}
  \item[$\bullet$] the \textit{1st Sample} is obtained by taking into consideration the data shared by two datasets: one from \citet{cap2013}, that is a sample of early-type galaxies, from which we take the mass of each galaxy with the respective  velocity dispersion, and the other from \citet{van2016}, from which we take the relative mass of the supermassive black hole. The sample is thus made up of 47 galaxies, excluding NGC4429, as its morphological classification is doubtful, and, as regards the NGC4486, we preferred to insert the most recent data obtained for the mass of its SMBH \citep{eve2019}.
      \vspace{5pt}
  \item[$\bullet$] The \textit{2nd Sample} is made up of 174 galaxies, obtained starting from the van den Bosch's sample \citep{van2016}, but excluding those galaxies whose mass of the SMBH shows a relative error on $\log(M_{\bullet}) $ greater than or equal to 1 \citep{bel2019}, NGC404, since its mass is too low for an SMBH and NGC4486b which ``deviates strongly from any correlation involving the mass of its black hole" \citet{sag2016}. Finally, we also preferred to exclude NGC221, NGC1277, NGC1316, NGC5845 and UGC1841.
      \vspace{5pt}
  \item[$\bullet$] The \textit{3rd Sample} consists only of the 108 early-type galaxies in the \textit{2nd Sample}.
      \vspace{5pt}
  \item[$\bullet$] The \textit{4th Sample} is composed of 71 objects, carried out from \citet{den2019}, from which we take the velocity dispersion and the mass of the SMBH for each galaxy, whereas  the mass of the corresponding galaxies $M_{Bu}$ comes from \citet{sag2016}.
      \vspace{5pt}
  \item[$\bullet$] The \textit{5th Sample} consists of all the data as given in \citet{sag2016}, choosing the same galaxies as in the previous sample.
      \vspace{5pt}
  \item[$\bullet$] The \textit{6th Sample} is obtained considering the sample discussed in Kormendy and Ho (2013), but reduced to $45$ galaxies after excluding the pseudobulges.

\end{itemize}

Both van den Bosch's samples do not explicitly contain the masses of the galaxies but include the effective radius $R_e$ of the host spheroidal component and the velocity dispersion $\sigma$ of the host galaxy. Then, the mass is computed as
\begin{equation}
M_{dyn}=\frac{5R_e \sigma^2}{G}
\end{equation}
where $G$ is the gravitational constant \citep{cap2006}.

\section{Model fitting and prediction}
The scaling relations we aim to fit to the data at hand are of the type

\begin{equation}
\label{mod1}
y=b + m x + \epsilon
\end{equation}
where $y$ is the response, $x$ the explanatory variable (or covariate), $\epsilon$ is the error term. According to an ordinary least squares (OLS) framework, it is assumed that the error term has null expected value and a constant variance that we denote by $\epsilon^2_{0}$. It is common to refer to $\epsilon_{0}$ as the intrinsic scatter.

The slope parameter $m$ is a direct measure of the strength of the linear association between the response $y$ and the covariate $x$. The OLS estimate of the slope parameter is given by
\begin{equation}
\hat m=\frac{s_{xy}}{s^2_x}=\rho_{xy}\frac{s_y}{s_x}
\end{equation}
where $s_{xy}$ is the covariance among $y$ and $x$, $s_x$ and $s_y$ are the standard deviation of $x$ and $y$, respectively, obtained as the square root of the corresponding variances $s^2_x$ and $s^2_y$. It can be noted that the estimate of the slope parameter can be expressed in terms of the sample Pearson linear correlation coefficient $\rho_{xy}=s_{xy}/s_xs_y$.
Let
\begin{equation}
	\label{mod1F}
	\hat \mu=\hat b + \hat m x
	\end{equation}
	be the fitted model. Then, the fitted values are $\hat \mu_i=\hat b + \hat m x_i$, $i=1,2,\ldots,n$. In expression (\ref{mod1F}) above, we use the symbol $\hat\mu$ rather than $\hat y$ to stress that, according to the hypothesis underlying model (\ref{mod1}), we are fitting the conditional expected value of the response variable given the value of the explanatory variable.
	
The goodness of fit is commonly assessed through the coefficient of determination $R^2\in[0,1]$.
It measures the share of total variability, as measured by the total sum of squares (also called deviance) $(n-1)s^2_y=\sum_{i=1}^n(y_i-\bar y)^2$, where $\bar y$ is the sample mean, of the response variable explained by the fitted regression model.
Actually, the total sum of squares can be decomposed as
$$
\sum_{i=1}^n(y_i-\bar y)^2=\sum_{i=1}^n (y_i-\hat\mu_i)^2+ \sum_{i=1}^n (\hat\mu_i-\bar y)^2 \ .
$$
Then,
\begin{equation}
R^2=1-\frac{\chi^2}{(n-1)s^2_y}
\end{equation}
where $\chi^2$ denotes the sum of squared errors
\begin{equation}
\chi^2=\sum_{i=1}^n \hat\epsilon_i^2=\sum_{i=1}^n (y_i-\hat\mu_i)^2= (n-2)\hat\epsilon_0^2,
\end{equation}
and $\hat\epsilon_0$ is the classical estimate of the intrinsic scatter. We find that $R^2=\rho_{xy}^2$.
In order to assess the validity of the fitted model, it is recommended to test the null hypothesis $H_0:m=0$ according to the statistic $t=\hat m/ se(\hat m)$, where $se(\hat m)=\frac{\hat\epsilon_0}{s_x\sqrt{n}}$ is the standard error and $t$ is distributed according to a Student's $t_{n-2}$ distribution. The model is valid when there is evidence supporting the importance of the covariate to explain part of the variability of the response, as measured by a small enough p-value $\mathrm{Prob(|t_{n-2}|>|t|)}$.

In addition to measure the quality of the fitted scaling relation in terms of goodness of fit, in this paper we also suggest to evaluate the prediction accuracy of the fitted model when the task is that of predicting new response values. Here, the prediction ability is fairly measured by cross-validation.
The basic idea behind cross-validation is to split the data into two subsets: one is used as a \textit{training sample}, i.e. to build the model, the other acts as a \textit{test sample} (or validation sample), to assess and validate the model. This validation is performed by comparing the test data with their predictions stemming from the model fitted on the train data. The splitting of the data is supposed to be executed randomly and repeated several times to get a fair estimate of the prediction accuracy of the fitted model.
Actually, cross-validation is a resampling method.

In this paper we used \textit{leave-one-out cross-validation}. The model is fitted $n$ times, every time omitting one observation. Then, for every fitted model, a prediction of the previously discarded observation is obtained as
$\hat\mu_{-i}=\hat b_{-i}+\hat m_{-i} x_i$, where the subscript $({-i})$ is used to stress the fact that estimation is based on the sample without the ith observation.
Prediction accuracy is measured by the prediction mean squared error
\begin{equation}
\epsilon_{pred}^2=\frac{1}{n}\sum_{i=1}^n(y_i-\hat\mu_{-i})^2=\frac{\chi^2_{pred}}{n}
\end{equation}

\section{Residuals Analysis and model comparison}
\label{res_an}
According to \citet{bar2017}, \citet{ber2007}, \citet{hop2007a,hop2007b} and \citet{sha2016,sha2019},
one way to establish the relative importance of one covariate $x$ with respect to another covariate $z$ in the linear relationship with the variable $y$, and check if the inclusion of the extra explanatory variable $z$ in the linear model (\ref{mod1}) is able to improve the fitted model, is to compare correlations among residuals. Correlations among residuals from scaling relations will be called conditional correlations. The technique works as follows:
\begin{enumerate}
		\item fit the regression of $y$ over $x$ and get residuals
		\begin{equation}r_{y|x}=y_i-(\hat b+ \hat m x_i);
        \end{equation}
		\item fit the regression of $z$ on $x$ according to the model $z= c+n x+\epsilon$ and get residuals
		\begin{equation}r_{z|x}=z_i-(\hat c+ \hat n x_i);
        \end{equation}
		\item consider the linear correlation between the residuals from the two fitted models, $r_{y|x}$ and $r_{z|x}$ respectively, that we denote $\rho_{yz|x}$;
     \vspace{5pt}
		\item repeat the procedure by exchanging the role of $x$ and $z$ and get $\rho_{yx|z}$;
     \vspace{5pt}
		\item compare conditional correlation coefficients: if $$\rho_{yx|z}>\rho_{yz|x}$$ then $x$ is more importantly correlated with $y$ relatively to $z$.
	\end{enumerate}

In order to give more insights on this type of residual analysis, it is worth to remark some statistical aspects that have been somewhat neglected in the papers cited above and may affect the discussion about the results.
Actually, a fair comparison between the conditional correlation coefficients should take into account the uncertainty that characterizes the estimates of conditional correlations driven by the fitted models. In other words, the comparison can not rely only on point estimates of conditional correlations, but on suitable statistical tests and, in a completely equivalent fashion, reporting confidence intervals.
Here, confidence intervals are obtained according to the classical Fisher Z test for linear correlation.

The comparison between conditional correlations does not only lead to state that one variable can be more important and fundamental in the scaling relation with respect to the other, but also to verify if the assumed model can be improved significantly in terms of goodness of fit by the inclusion of the extra variable. Assume that we fit the scaling relation in (\ref{mod1}). Actually, when zero is outside the $(1-\alpha)$ confidence interval around $\rho_{yz|x}$, this means that the inclusion of $z$ leads to a significant reduction in the residual sum of squared errors (at the level $\alpha$) and, hence, to an improvement of the model in terms of goodness of fit as measured by $R^2$ and, likely, in predictive accuracy.

Let us assume that $z$ is also important. Omitting $z$ implies a bias in the estimate of the slope parameter that is equal to the fitted slope of the regression of $z$ on $x$, that is $\rho_{xz}\frac{s_z}{s_x}$.  Then, the larger the correlation, the larger the omitted variable bias. Table \ref{tab1} gives the linear correlation coefficient between kinetic energy and velocity dispersion (both on a log scale) for each Sample under study. A strong correlation emerges between $x$ and $z$.

Since,  for  van den Bosch  sample, the kinetic energy is
\begin{equation}
KE \propto M_{dyn} \sigma^2 \propto (R_e \sigma^2) \sigma^2,
\end{equation}
then a strong correlation between the effective radius of host spheroidal component Re and the velocity dispersion exists. If this relation is linear,
the slopes of the $M_{\bullet}-\sigma$ and $M_{\bullet}-M_{G}\sigma^2$ relations should differ by a factor of 5. However, this is not always tha case. Actually, we verified with the second sample containing 174 galaxies, that the fitted slope of the $R_{e}-\sigma$ relation is $0.63\pm0.13$, that differs significantly from the slope equal to $1$ found in the work of \citet{ber2003}.

\begin{table}[ht]
		\caption{Linear correlation between kinetic energy and velocity dispersion (both on a log scale) for each Sample}
	\begin{center}
		\smallskip
		\begin{tabular}{@{}rccc@{}}
			\tableline
			\tableline
			Sample&&$\rho_{xz}$&\\
			\tableline
			\textsl{Cappellari}&&0.912&\\
			\textsl{van den Bosch\_174}&&0.939&\\
			\textsl{van den Bosch\_108}&&0.945&\\
			\textsl{de Nicola-Saglia}&&0.937&\\
			\textsl{Saglia}&&0.937&\\
			\textsl{Kormendy-Ho}&&0.948&\\
			\tableline
		\end{tabular}
	\end{center}

	\label{tab1}
\end{table}

The conditional correlation $\rho_{yz|x}$ is proportional to the estimate of the slope of the regression of the residuals from step 1 over the residuals from step 2 above. Let denote this estimate as $\hat m_{yz|x}$. It can be proved that  $\hat m_{yz|x}=\hat m_z$ where $m_z$ is the coefficient of the model
\begin{equation}\label{uno}
y=b+m_x x+ m_z z + \epsilon \
\end{equation}
and $\hat m_z$ the corresponding least squares estimate.
This result follows from the Frisch-Waugh theorem \citep{sw:2015}. The reader is also pointed to \cite{sha2016} and the interesting Appendix B in \cite{sha2017}.

The inclusion of an extra covariate $z$ always lead to an improvement in terms of goodness of fit, in the sense that $R^2$ increases and the residual sum of squared errors becomes smaller. However, one is expected to check if this reduction is large enough to assess the importance of $z$. According to the residual analysis described so far, the reduction is significant when the confidence interval around $\rho_{yz|x}$ does not include zero. In an equivalent fashion, one could fit the {\it complete} model in (\ref{uno}) and check the significance of the test of nullity about the slope coefficient $m_z$. To this end, one should use a classical Students' t test with $n-3$ degrees of freedom, denoted by $t_{n-3}$.

In a completely equivalent fashion, the relative importance of one explanatory variable with respect to the other can be directly assessed by measuring the reduction in the residual sum of squared errors. In other words, we are interested in the difference
\begin{equation}
\Delta=(\chi^2-\chi^2_{complete}),
\end{equation}
where $\chi^2$ is the sum of squared errors from the fitted model with only $x$ or $z$ and $\chi^2_{complete}$ is the sum of squares from model (\ref{uno}).
In order to verify if the reduction $\Delta$ is large enough to assess the importance of the extra explanatory variable, we consider the test statistic
\begin{equation}
F=\frac{\Delta}{\hat\epsilon^2_{complete}}
\end{equation}
where $\hat\epsilon^2_{complete}=\frac{\chi^2_{complete}}{n-3}$. The p-value
\begin{equation}
\mathrm{Prob}(F_{1,n-3}>\Delta)
\end{equation}
is a direct measure of evidence, where $F_{1,n-3}$ is a Fisher random variable with one and $n-3$ degrees of freedom. The test is significant when the p-value is small enough. The equivalence between this approach and the strategy based on test the hypothesis $H_0: m_z=0$ (or $H_0: m_x=0$) in model (\ref{uno}) can be remarked by noting that $F_{1,n-3}=t^2_{n-3}$.

To sum up, comparing the relative importance of the variable $x$ with respect to the variable $z$ results in fitting the complete model (\ref{uno}). The comparison can be pursued in an equivalent fashion according to inferences (tests or confidence intervals) on:
\begin{enumerate}
	\item the coefficients in the complete model;
     \vspace{5pt}
	\item the reduction in the residual sum of squared errors;
     \vspace{5pt}
	\item conditional correlations.
	\end{enumerate}

\section{Fitted scaling relations: empirical results}
\label{results1}
In this section we present and discuss the results we obtained for each sample introduced in Sect. \ref{sample} for what concerns goodness of fit and prediction accuracy of the scaling relations $M_{\bullet}- \sigma$ and $M_{\bullet}- M_{G}\sigma^2$. To be precise, both the response and the covariate are on a logarithmic scale.
The results are based on {\tt R} routines and codes, whereas in \citet{ian2020} the OLS fit was obtained from Mathematica and other fits were obtained from the LINMIX\_ERR \citep{kel2007} and MPFITEXY \citep{tre2002} routines. The latter techniques have not been considered here, but prediction accuracy could have been evaluated by leave-one-out cross-validation as well.

As a general comment, the $M_{\bullet} - M_{G}\sigma^2$ scaling relation performs quite satisfactory when compared to the classical one. The entries in Table \ref{fit6} allow the comparison among the considered scaling relations, both in terms of goodness of fit and prediction error. The former can be assessed looking at the value of $R^2$ or, in an equivalent fashion, $\hat\epsilon_0$ and the corresponding $\chi^2$. The latter through the value $\epsilon_{pred}$ and the corresponding $\chi^2_{pred}$. A general advise is to prefer the model returning the smallest fitted intrinsic scatter (that means larger $R^2$) and prediction error. In this respect, there could be evidence supporting the validity of both fitted models but one can still provide (even slight) superior goodness of fit and prediction accuracy. As one referees pointed out, it worth noting that the sum of squared (prediction) errors are not comparable across samples, whereas this is true for the fitted intrinsic scatter or the prediction error.

\begin{table*}[!h]
	\caption{Fitted scaling relations, goodness of fit and prediction accuracy}
	
	\begin{center}
		
		\begin{tabular}{rccccccc}
			\tableline
			\tableline
			Scaling relation&$\hat b \pm se(\hat b)$ & $\hat m \pm se(\hat m)$ &$R^2$&$\hat\epsilon_0$&$\chi^2$&$\epsilon_p$&$\chi^2_p$\\
			\tableline
			\noalign{\smallskip}
			\multicolumn{8}{c}{\textsl{\textbf{1st Sample: Cappellari}}} \\
			\noalign{\smallskip}
			\vspace*{4pt}\textsl{$Log\left(M_{\bullet}\right)-$ $Log\left(\displaystyle M_{JAM}\sigma^2 \over \displaystyle c^2 \right) $} &3.88$\pm 0.41$& 0.96$\pm 0.09$&  $0.715$ &0.413& $7.68$ & $0.419$ & $8.27$ \\
			\vspace*{4pt}\textsl{$Log\left(M_{\bullet}\right)-Log\left(\sigma\right)$} & -2.75$\pm 1.03$& 4.88$\pm 0.46$&    0.714  &$0.414$&
			$7.69$ & $0.427$ & 8.56 \\
			\tableline
			\noalign{\smallskip}
			\multicolumn{8}{c}{\textsl{\textbf{2nd Sample: van den Bosch\_174}}} \\
			\noalign{\smallskip}
			
			\vspace*{4pt}\textsl{$Log\left(M_{\bullet}\right)-$ $Log\left(\displaystyle M_{dyn}\sigma^2 \over \displaystyle c^2\right) $} &3.52$\pm 0.26$& 0.95$\pm 0.05$&  $0.652$ &0.569& $55.70$ & $0.572$ & $56.85$ \\
			\vspace*{4pt}\textsl{$Log\left(M_{\bullet}\right)-Log\left(\sigma\right)$} & -2.91$\pm 0.55$& 4.89$\pm 0.24$&    0.702  &$0.526$&
			$47.64$ & $0.529$ & 48.77 \\
			\tableline
			\noalign{\smallskip}
			\multicolumn{8}{c}{\textsl{\textbf{3rd Sample: van den Bosch\_108}}}\\
		
			\noalign{\smallskip}
			
			\vspace*{4pt}\textsl{$Log\left(M_{\bullet}\right)-$ $Log\left(\displaystyle M_{dyn}\sigma^2 \over \displaystyle c^2\right) $} &4.17$\pm 0.24$& 0.86$\pm 0.05$&  $0.758$ &0.423& $18.96$ & $0.427$ & $19.65$ \\
			\vspace*{4pt}\textsl{$Log\left(M_{\bullet}\right)-Log\left(\sigma\right)$} & -2.44$\pm 0.59$& 4.73$\pm 0.25$&    0.766  &$0.416$&
			$18.35$ & $0.421$ & 19.12 \\

			\tableline
			
			\noalign{\smallskip}
			\multicolumn{8}{c}{\textsl{\textbf{4th  sample: De Nicola - Saglia}}} \vspace*{2pt}\\
			\noalign{\smallskip}
			
			\vspace*{4pt}\textsl{$Log\left(M_{\bullet}\right)-$ $Log\left(\displaystyle M_{Bu}\sigma^2 \over \displaystyle c^2\right) $} &5.19$\pm 0.17$& 0.72$\pm 0.04$&  $0.846$ &0.387& $10.31$ & $0.392$ & $10.89$ \\
			\vspace*{4pt}\textsl{$Log\left(M_{\bullet}\right)-Log\left(\sigma\right)$} & -2.94$\pm 0.62$& 4.92$\pm 0.27$&    0.830  &$0.406$&
			$11.35$ & $0.413$ & 12.11 \\
			\tableline

			\noalign{\smallskip}
			\multicolumn{8}{c}{\textsl{\textbf{5th Sample: Saglia}}}\\
			\noalign{\smallskip}
			
			\vspace*{4pt}\textsl{$Log\left(M_{\bullet}\right)-$ $Log\left(\displaystyle M_{Bu}\sigma^2 \over \displaystyle c^2\right) $} &5.17$\pm 0.17$& 0.72$\pm 0.04$&  $0.845$ &0.392& $10.60$ & $0.398$ & $11.22$ \\
			\vspace*{4pt}\textsl{$Log\left(M_{\bullet}\right)-Log\left(\sigma\right)$} & -3.05$\pm 0.62$& 4.97$\pm 0.27$&    0.832  &$0.408$&
			$11.46$ & $0.416$ & 12.26 \\
			\tableline
			\noalign{\smallskip}
			\multicolumn{8}{c}{\textsl{\textbf{6th Sample: Kormendy - Ho}}} \\
			\noalign{\smallskip}
			
			\vspace*{4pt}\textsl{$Log\left(M_{\bullet}\right)-$ $Log\left(\displaystyle M_{Bu}\sigma^2 \over \displaystyle c^2 \right) $} &4.97$\pm 0.20$& 0.78$\pm 0.04$&  $0.886$ &0.274& $3.23$ & $0.278$ & $3.48$ \\
			\vspace*{4pt}\textsl{$Log\left(M_{\bullet}\right)-Log\left(\sigma\right)$} & -1.39$\pm 0.68$& 4.29$\pm 0.29$&    0.837  &$0.328$&
			$4.62$ & $0.339$ & 5.16 \\
			\tableline
		\end{tabular}
	\end{center}

	\label{fit6}

\end{table*}

In details, these are the results for each of the six considered samples.
\begin{itemize}
	\item[$\bullet$] \textit{Cappellari}. The entries in Table \ref{fit6} state that the $M_{\bullet}-M_{JAM}\sigma^2$ relation exhibits a quite comparable behavior with the $M_{\bullet}-\sigma$ both in terms of intrinsic error (goodness of fit) and prediction error. The difference in terms of fitted intrinsic scatter (and $R^2$) is negligible whereas the former scaling relation exhibits a lower prediction error than the latter.
     \vspace{5pt}
\item[$\bullet$] \textit{van den Bosch\_174}.
In this sample, the situation is different from the previous one. The $M_{\bullet}-\sigma$ relation outperforms the $M_{\bullet}-M_{dyn}\sigma^2$ relation both in terms of goodness of fit and prediction error with a $70\%$ rate of explained variability, whereas it is $65\%$ for the latter.
     \vspace{5pt}
\item[$\bullet$] \textit{van den Bosch\_108}.
From Table \ref{fit6}, the $M_{\bullet}-\sigma$ relation shows slightly better performances in terms of goodness of fit and prediction accuracy.
     \vspace{5pt}
\item[$\bullet$] \textit{de Nicola - Saglia}.
The $M_{\bullet}-M_{Bu}\sigma^2$ relation performs better than the $M_{\bullet}-\sigma$ scaling relation both in terms of goodness of fit and prediction error: explained variability of the former is about $85\%$ compared to $83\%$ and the difference in the sum of squared prediction errors is not negligible.
     \vspace{5pt}
	\item[$\bullet$] \textit{Saglia}.
From Table \ref{fit6}, we observe results similar to the above case. The $M_{\bullet}-M_{Bu}\sigma^2$ relation still highlights a superior goodness of fit and prediction capability.
     \vspace{5pt}
	\item[$\bullet$] \textit{Kormendy-Ho}. According to the fitted models shown in Table \ref{fit6}, the $M_{\bullet}-M_{Bu}\sigma^2$ relation here outperforms the $M_{\bullet}-\sigma$ relation, since it
	exhibits a lower intrinsic scatter with $89\%$ of explained variability compared to $84\%$ and a remarkable lower prediction error.
	
\end{itemize}

\begin{figure*}[!h]
\centering%
\subfigure[]
{\includegraphics[width=79mm]{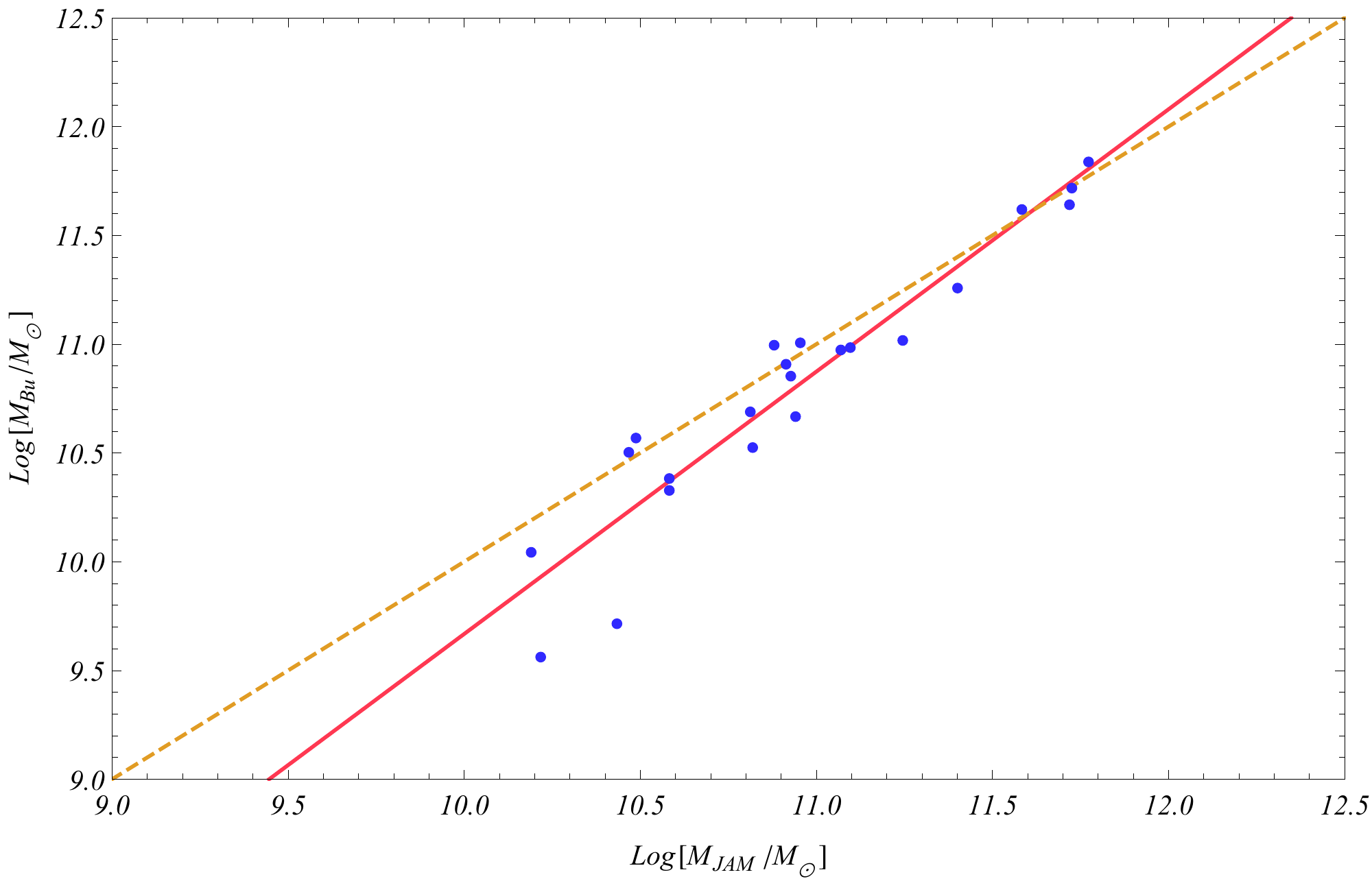}}\qquad\qquad
\subfigure[]
{\includegraphics[width=79mm]{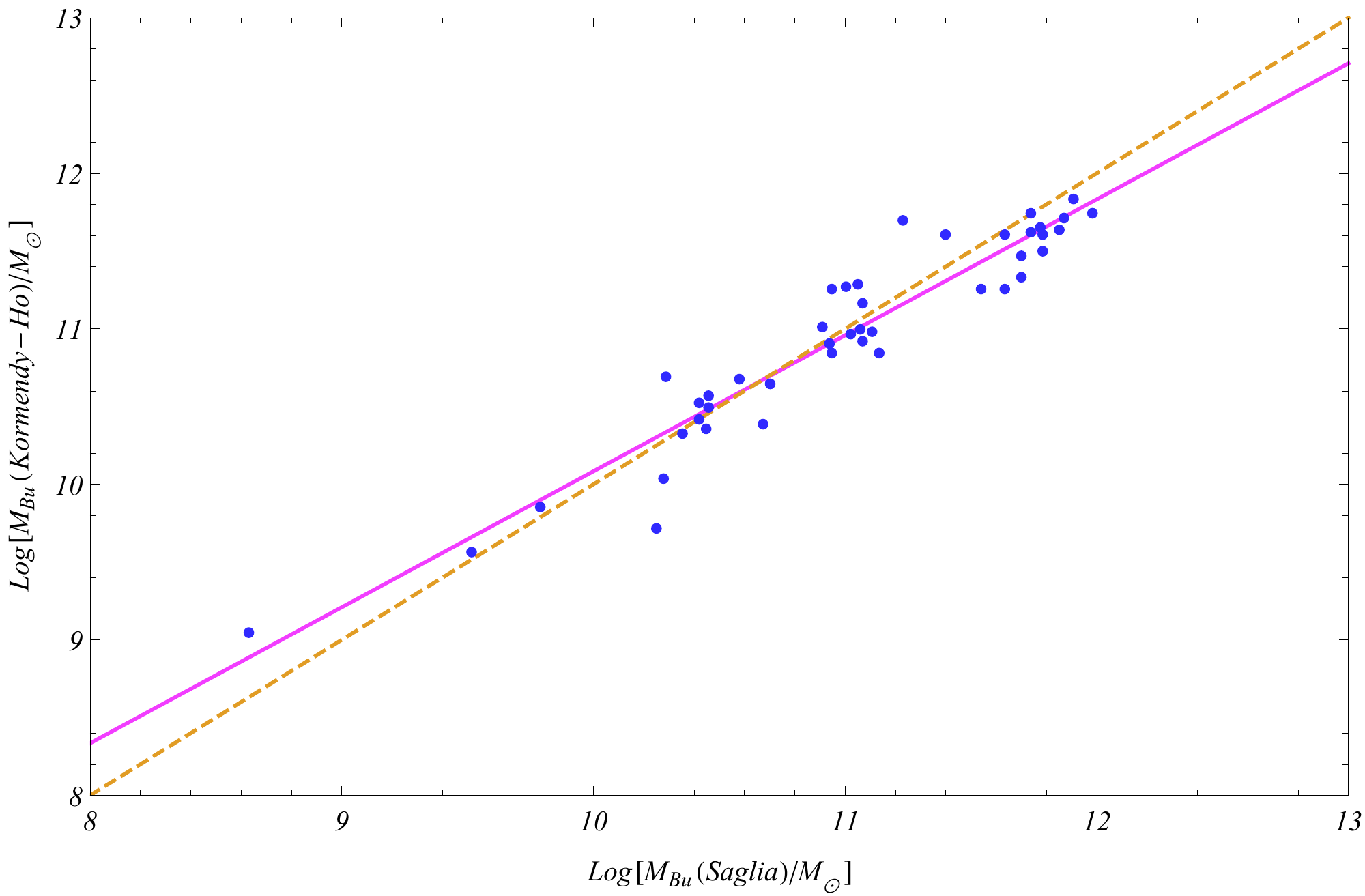}}\qquad\qquad
\subfigure[]
{\includegraphics[width=79mm]{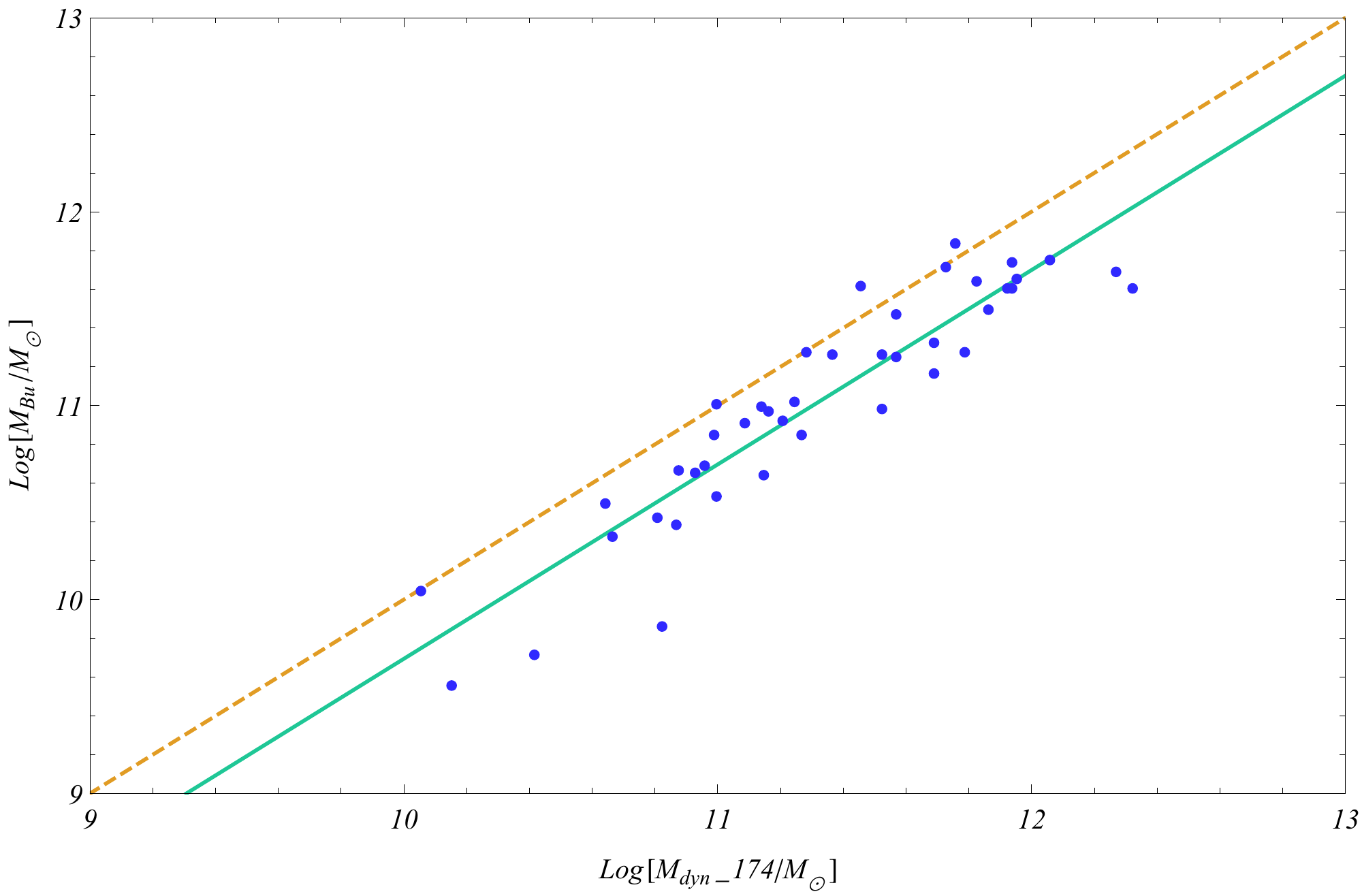}}\qquad\qquad
\subfigure[]
{\includegraphics[width=79mm]{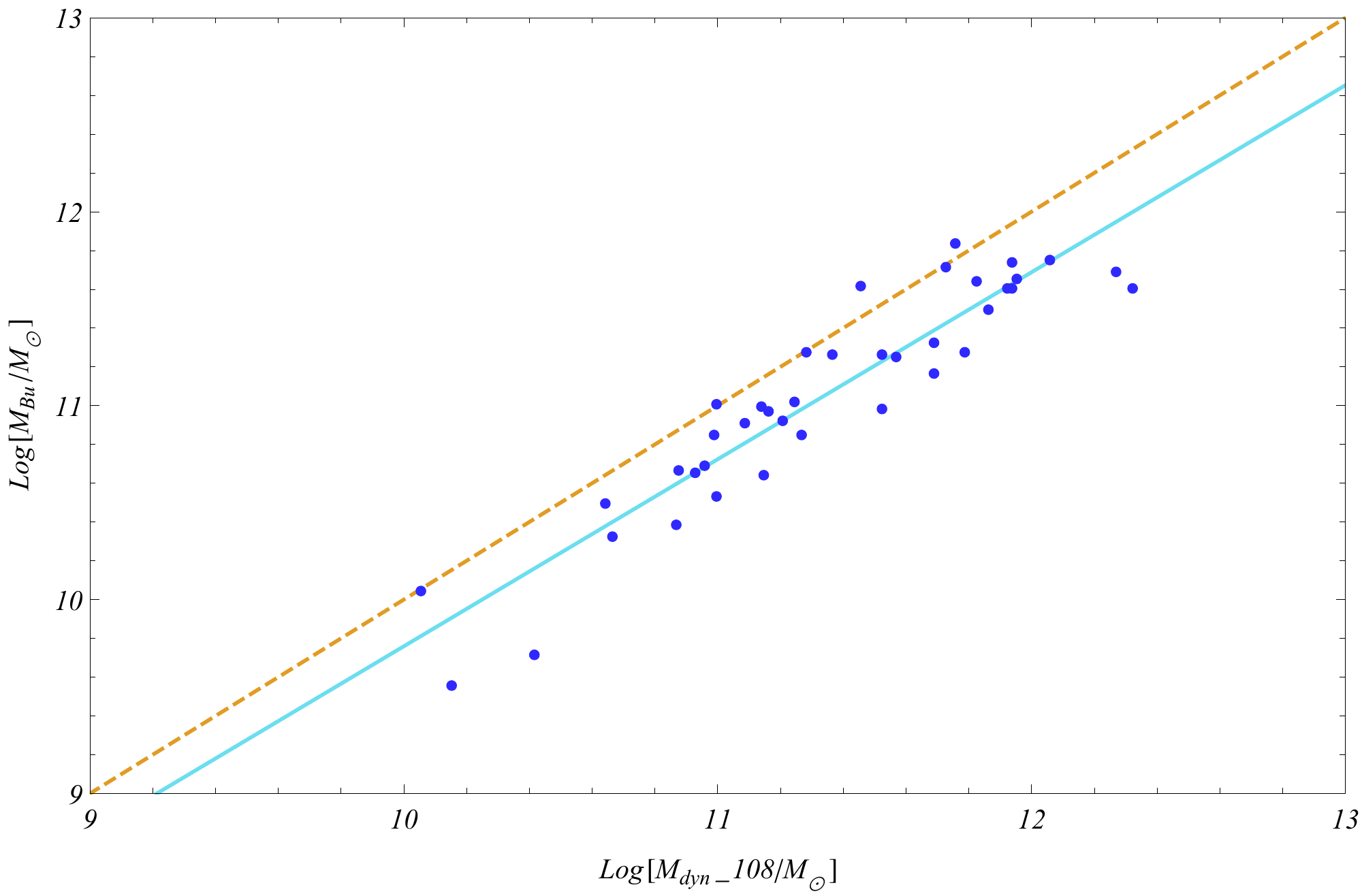}}
\caption{Comparison between masses from different samples. OLS fitted line (solid line) and expected line (dashed line) if the masses of both samples had the same values.
Only those galaxies common to both samples have been used.	
		\newline
(a) KormendyHo-Cappellari with OLS fitted line $Log\> M_{Bu}=(-2.39 \pm 0.99)+(1.21\pm 0.09)\> Log\> M_{JAM}$; \newline
(b) KormendyHo-Saglia with OLS fitted line $Log\> M_{Bu}(K-H)=(1.34 \pm 0.48)+(0.87\pm 0.04)\> Log\> M_{Bu}(S)$; \newline
(c) KormendyHo-van den Bosch\_174 with OLS fitted line $Log\> M_{Bu}=(-0.33 \pm 0.77)+(1.00\pm 0.07)\> Log\> M_{dyn}\_174$ and \newline
(d) KormendyHo-van den Bosch\_108 with OLS fitted line $Log \>M_{Bu}=(0.12 \pm 0.72)+(0.96\pm 0.06)\> Log\> M_{dyn}\_108$. }
\label{fig:1}
\end{figure*}

It is worth noting that the differeces in the fitted slopes found in the Cappellari and van den Bosch samples on the one hand, and Saglia sample on the other, is mainly due to a different estimate of the masses, as it has already been demonstrated and motivated in the paper by \citet{ian2020}, according to the corresponding figures therein. In order to complete the discussion made in the previous paper, we explicitly carried out the same analysis for the  Kormendy-Ho sample, not included in the previous work. From Fig. \ref{fig:1}, we notice that the trend of the masses of Kormendy-Ho is, on average, proportional  to that of van den Bosch,  but they are also very close to the set of masses from the Saglia sample. Then, the estimate of  the slope of $M_{\bullet}-M_{Bu}\sigma^2$ from the Kormendy-Ho sample is in the middle between the fitted slopes in the third and fourth samples. We learn that, in order to draw definitive conclusions about the slope of the relation  $M_{\bullet}-M_{Bu}\sigma^2$, it is crucial to choose with great accuracy the set of masses of galaxies to refer to.

\section{Residual analysis}
\label{results2}
In Sect. \ref{res_an} we focused on the role of residual analysis as discussed in \cite{bar2017} and \cite{sha2019} and highlighted its fundamental statistical aspects.

In  \cite{bar2017} and \cite{sha2019} the interest was in
studying correlations between the residuals from various scaling relations as an efficient way of comparing the relative importance of velocity dispersion and the mass of the galaxy $M_G$. Along the same line, here we consider the role played by $M_{G}\sigma^2$ and velocity dispersion. In light of the comments in Sect. \ref{res_an}, we first prefer to discuss again the results from \cite{sha2019} in Subsect. \ref{a1} and then to investigate the relative importance of the kinetic energy, characterizing the scaling relation $M_{\bullet} - M_{G}\sigma^2$ \citep{feo2005, feo2007, feo2009, feo2014}, compared to the velocity dispersion $\sigma$ in Subsect. \ref{a2}.

\subsection{\textit{Mass of the galaxy and velocity dispersion}}
\label{a1}
The entries  in Table \ref{cor7} give conditional correlations along with $95\%$ confidence intervals. Let $y=M_{\bullet}$,  $z=\sigma$, $w=M_G$.
In all considered samples, but the first van de Bosch sample, the complete model, with both variables $z$ and $w$ improves over the model with only one of them. In the first van de Bosch's sample, the velocity dispersion alone suffices to explain the SMBH. As a general result, velocity dispersion always shows a larger relative importance with respect to the mass of the galaxy, even if evidence of such superiority (at a $5\%$ significance level) can be found only in the first van de Bosch's sample.

\subsection{\textit{Kinetic energy and velocity dispersion}}
\label{a2}
The results concerning the analysis of correlations among residuals from scaling relations when the kinetic energy is considered in place of the mass of the galaxy are presented in Table \ref{cor7}, whereas those about the reduction in the sum of squared errors are in Table \ref{F6}.

\begin{figure*}[!h]
\centering%
\subfigure[]
{\includegraphics[scale=0.22]{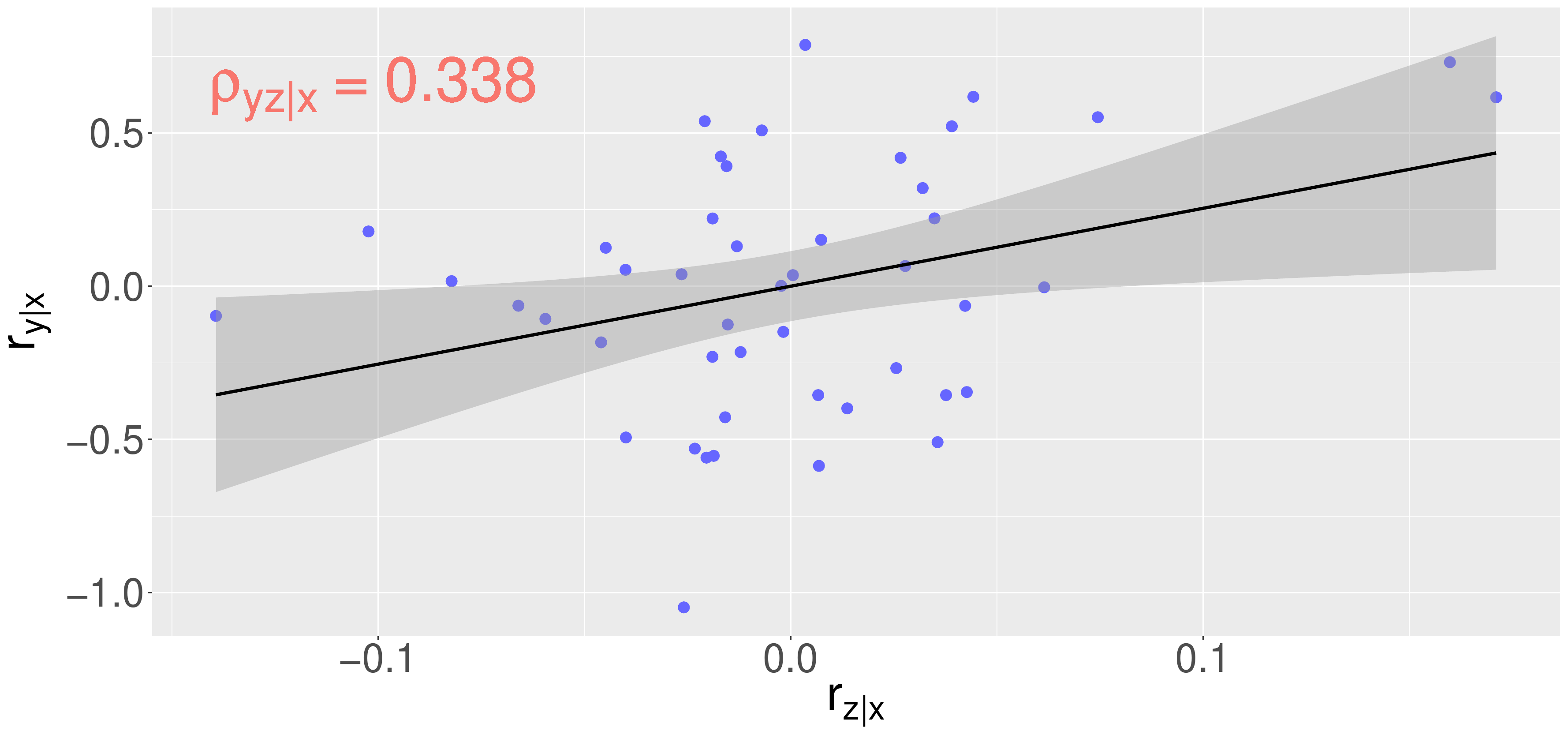}}\qquad\qquad
\subfigure[]
{\includegraphics[scale=0.22]{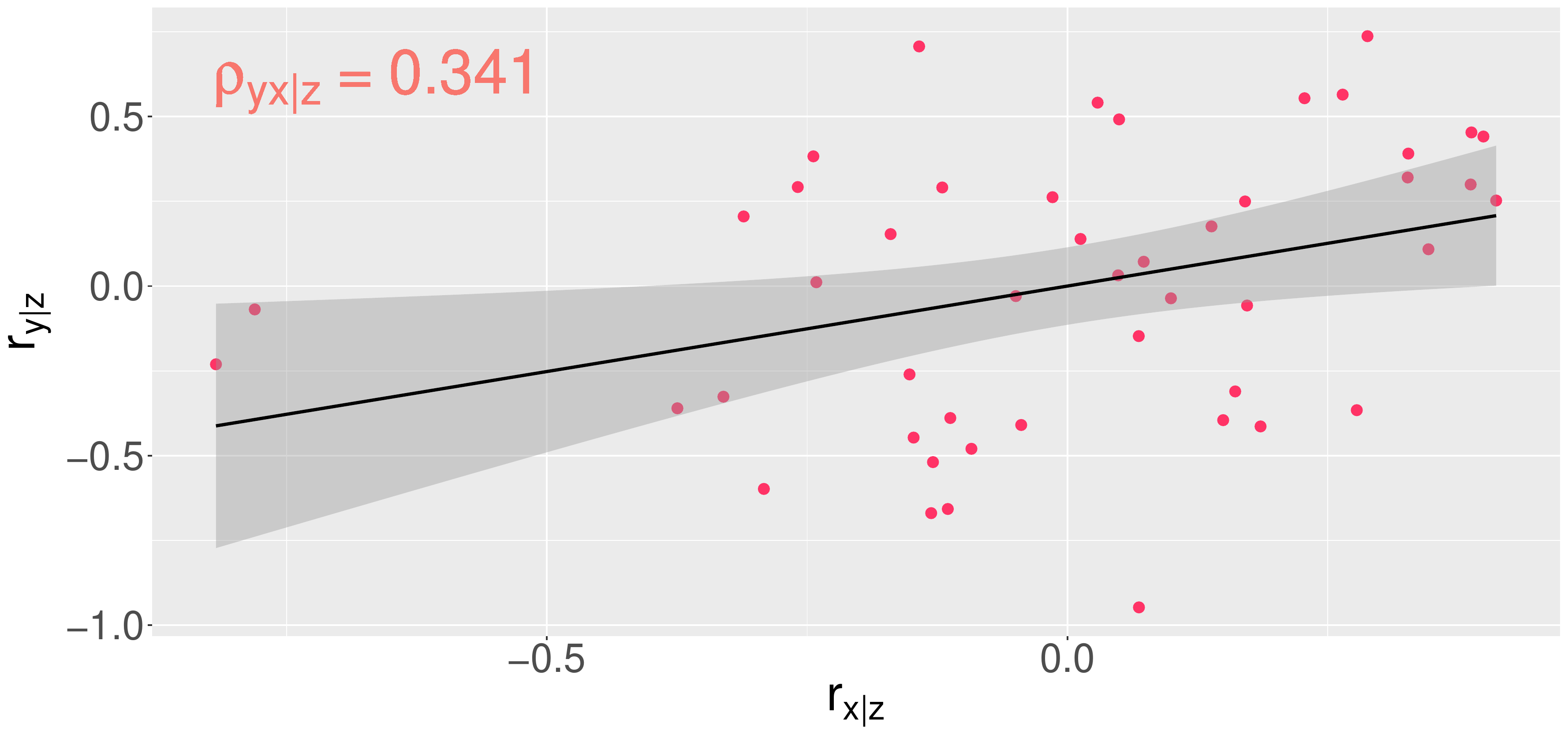}}\qquad\qquad
\subfigure[]
{\includegraphics[scale=0.22]{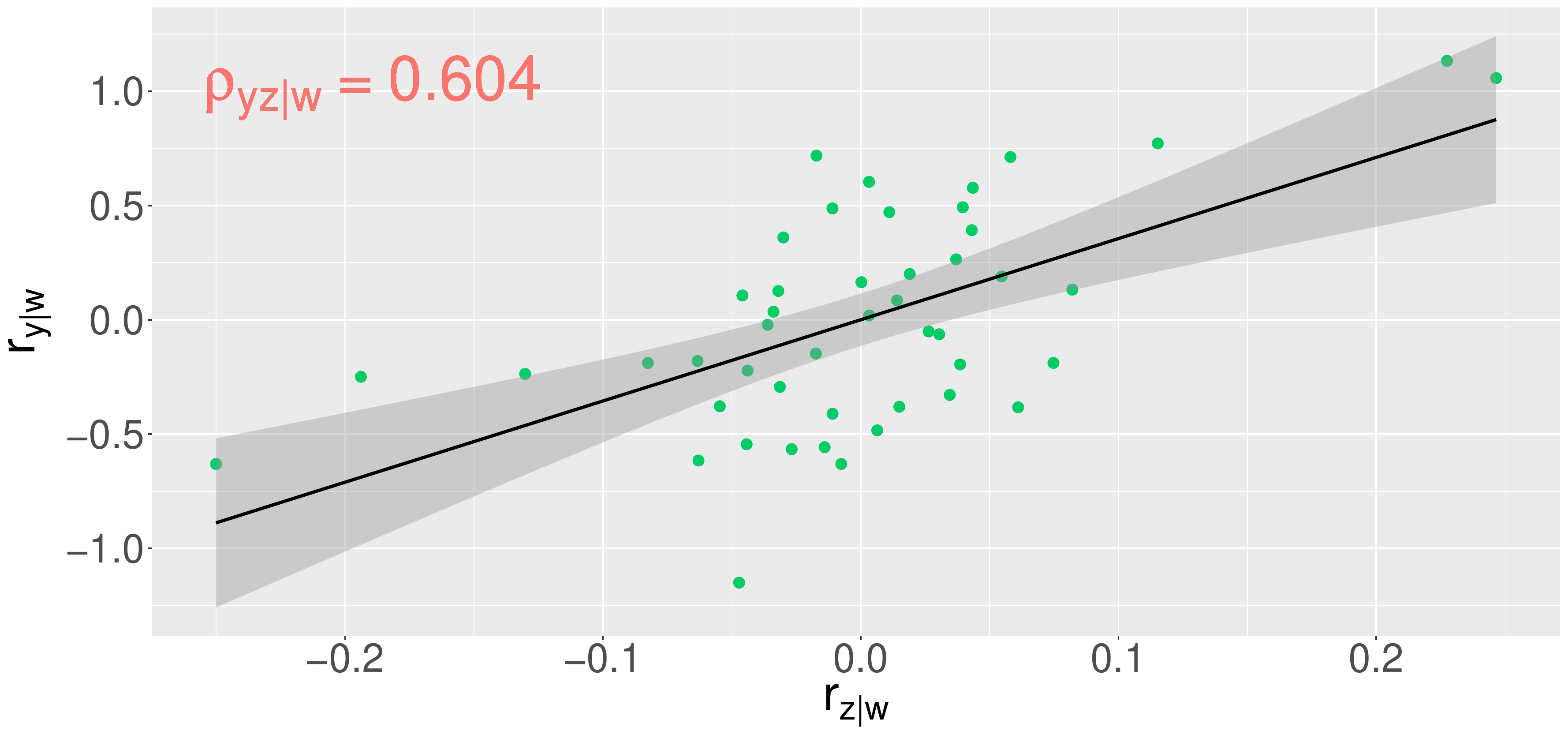}}
\caption{Correlation between residuals  for Cappellari's Sample, with $y=M_{\bullet}$,  $x=M_{G}\sigma^2$, $z=\sigma$, $w=M_G$: (a) $\rho_{yz|x}$, (b) $\rho_{yx|z}$ and (c) $\rho_{yz|w}$. }
\label{fig:cap}
\end{figure*}

\begin{figure*}[!h]
\centering%
\subfigure[]
{\includegraphics[scale=0.22]{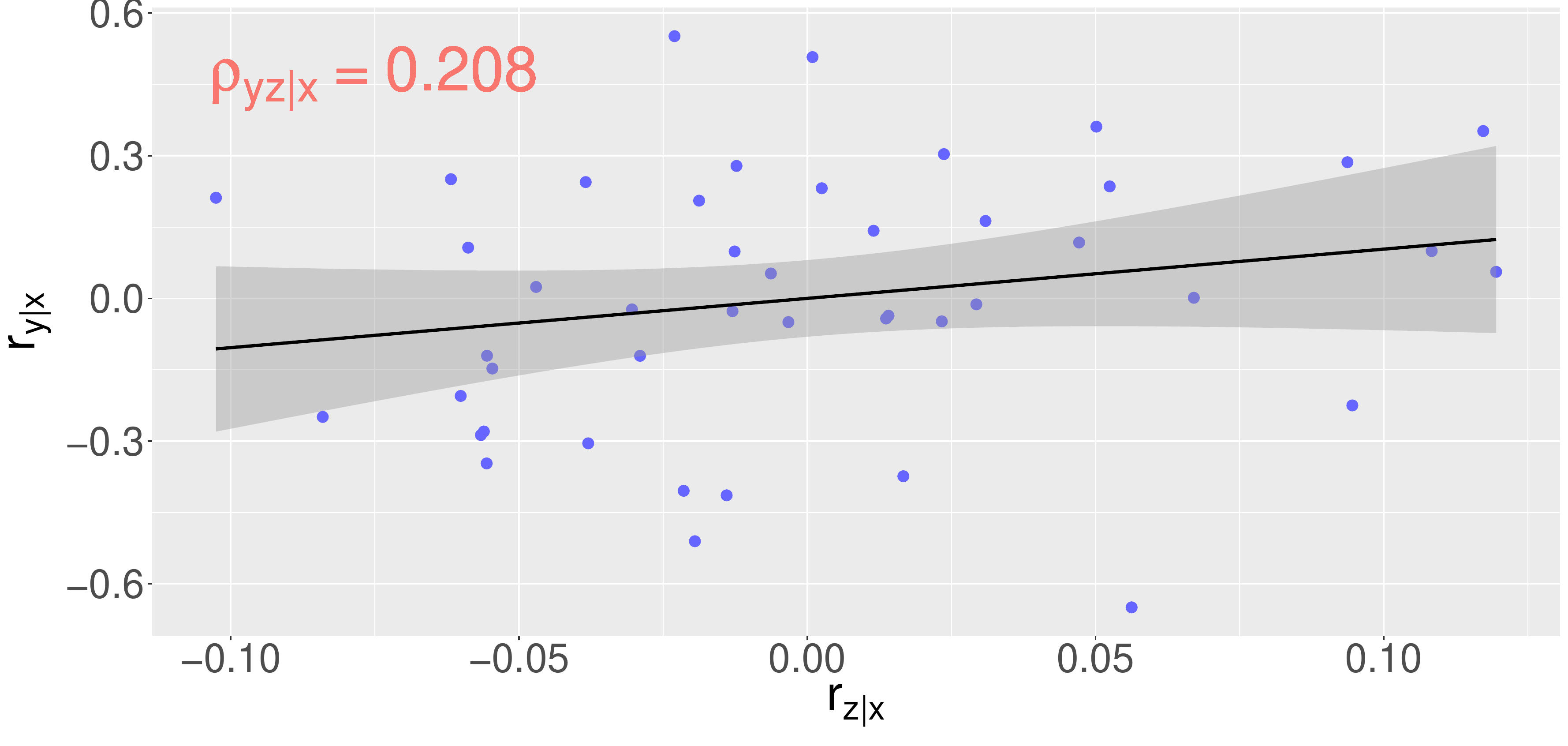}}\qquad\qquad
\subfigure[]
{\includegraphics[scale=0.22]{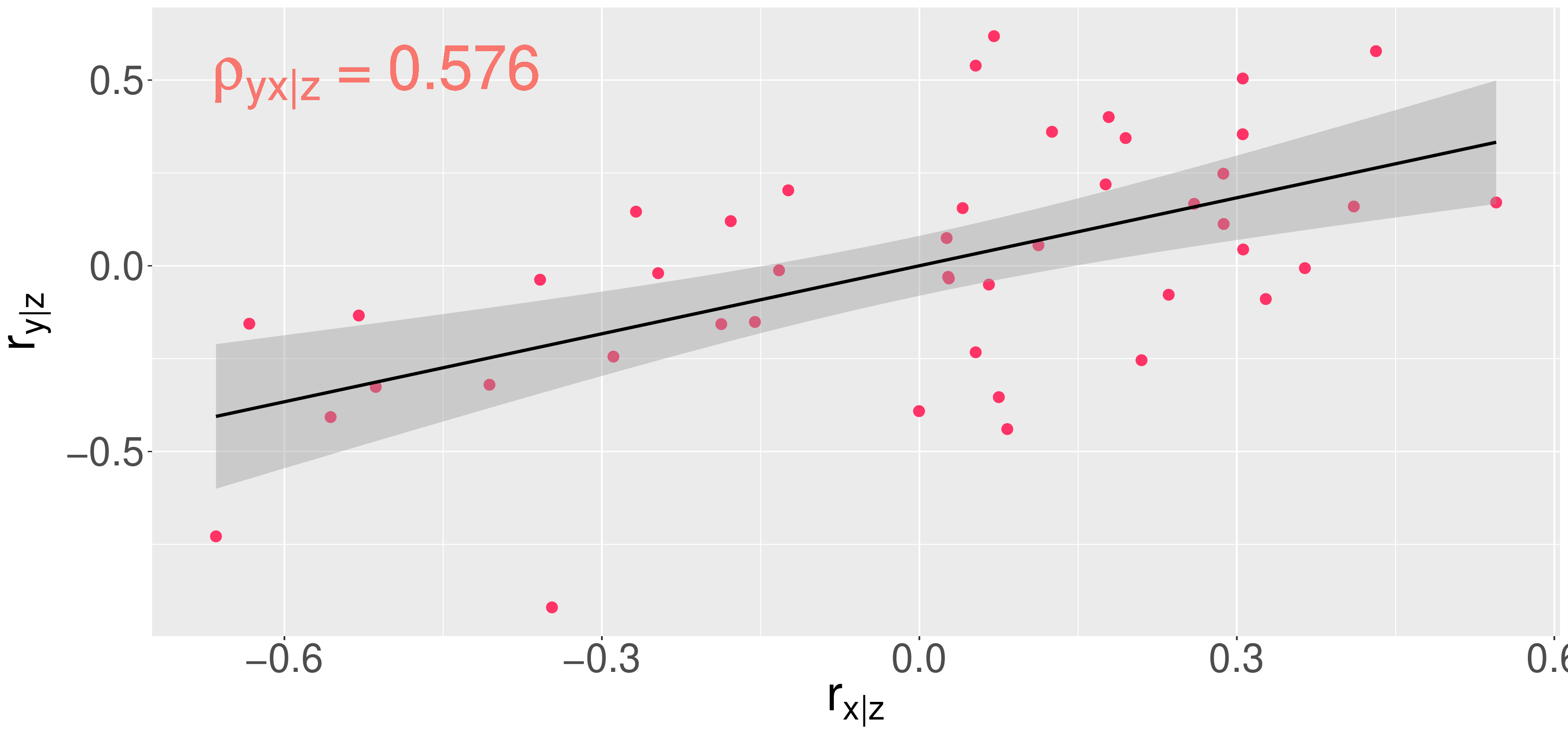}}\qquad\qquad
\subfigure[]
{\includegraphics[scale=0.22]{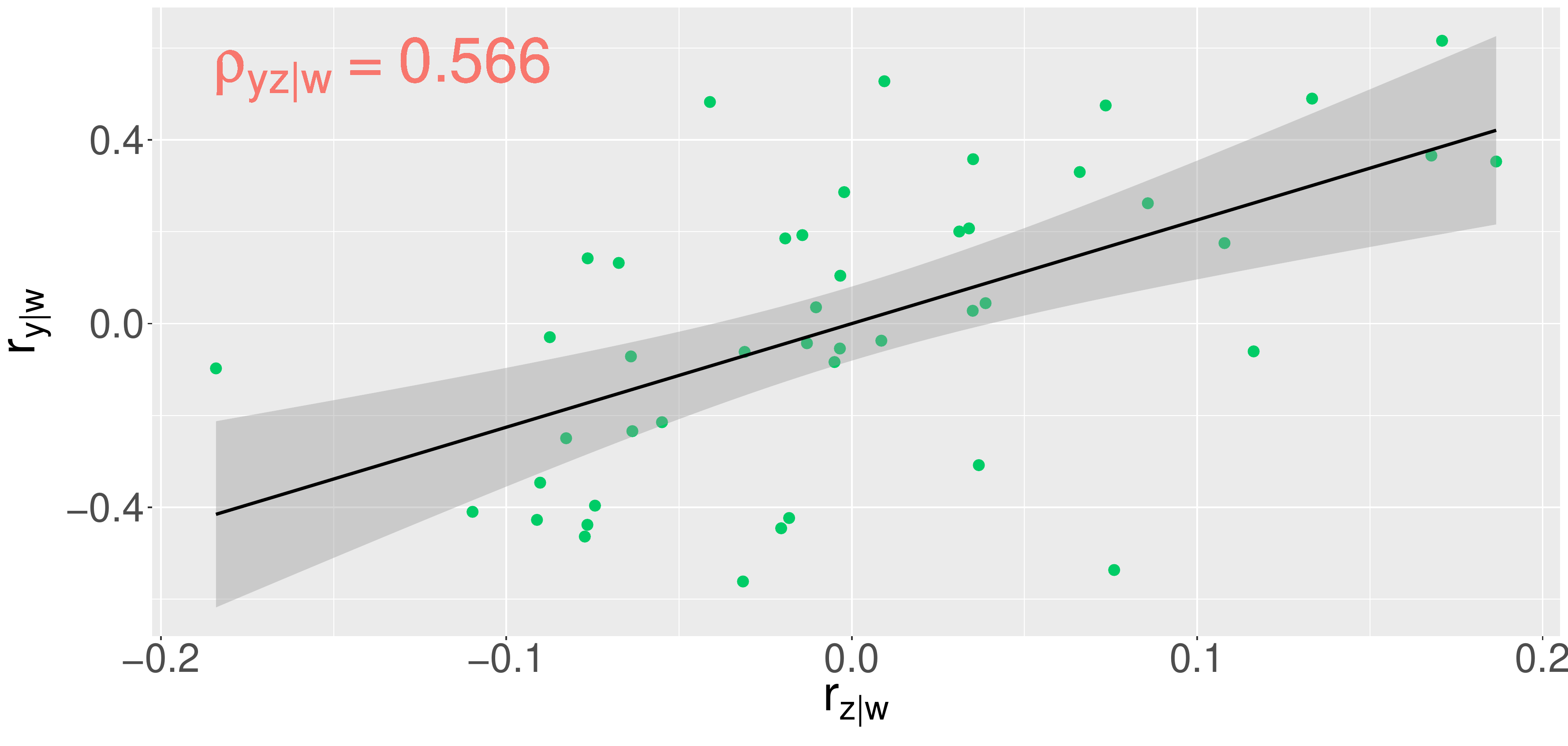}}
\caption{Correlation between residuals for the Kormendy-Ho's Sample,with $y=M_{\bullet}$,  $x=M_{G}\sigma^2$, $z=\sigma$, $w=M_G$: (a) $\rho_{yz|x}$, (b) $\rho_{yx|z}$ and (c) $\rho_{yz|w}$.}
\label{fig:kh}
\end{figure*}

There is no evidence of any significant difference between the magnitude of the correlations coefficients for all the samples except for the first van de Bosch's sample. Then, with the exception of the latter sample, we notice that, the importance of kinetic energy is larger, even if not significant, than that of velocity dispersion. In contrast, in Subsect. \ref{a1} we stated that there was no ambiguity about the preeminent role of velocity dispersion over the mass of the galaxy. Moreover, the introduction of the kinetic energy in addition to dispersion velocity improves goodness of fit since it leads to a significant reduction in the sum of squared errors, still with the exception of the first van de Bosch's sample.
The conditional correlations $\rho_{yz|w}$ are always larger than the conditional correlations $\rho_{yz|x}$, where $x=M_G \sigma^2/c^2$. This means that kinetic energy has larger predictive power than the mass of the galaxy.
It is worth noting that if $z=\sigma$ is fixed, $x$ and $w=M_G$ are statistically the same. As a result, $\rho_{yw|z}=\rho_{yx|z}$ and this leads us to insert only three subfigures for each figure (Figs. \ref{fig:cap} $\div$ \ref{fig:kh}).

\begin{figure*}[!h]
\centering%
\subfigure[]
{\includegraphics[scale=0.22]{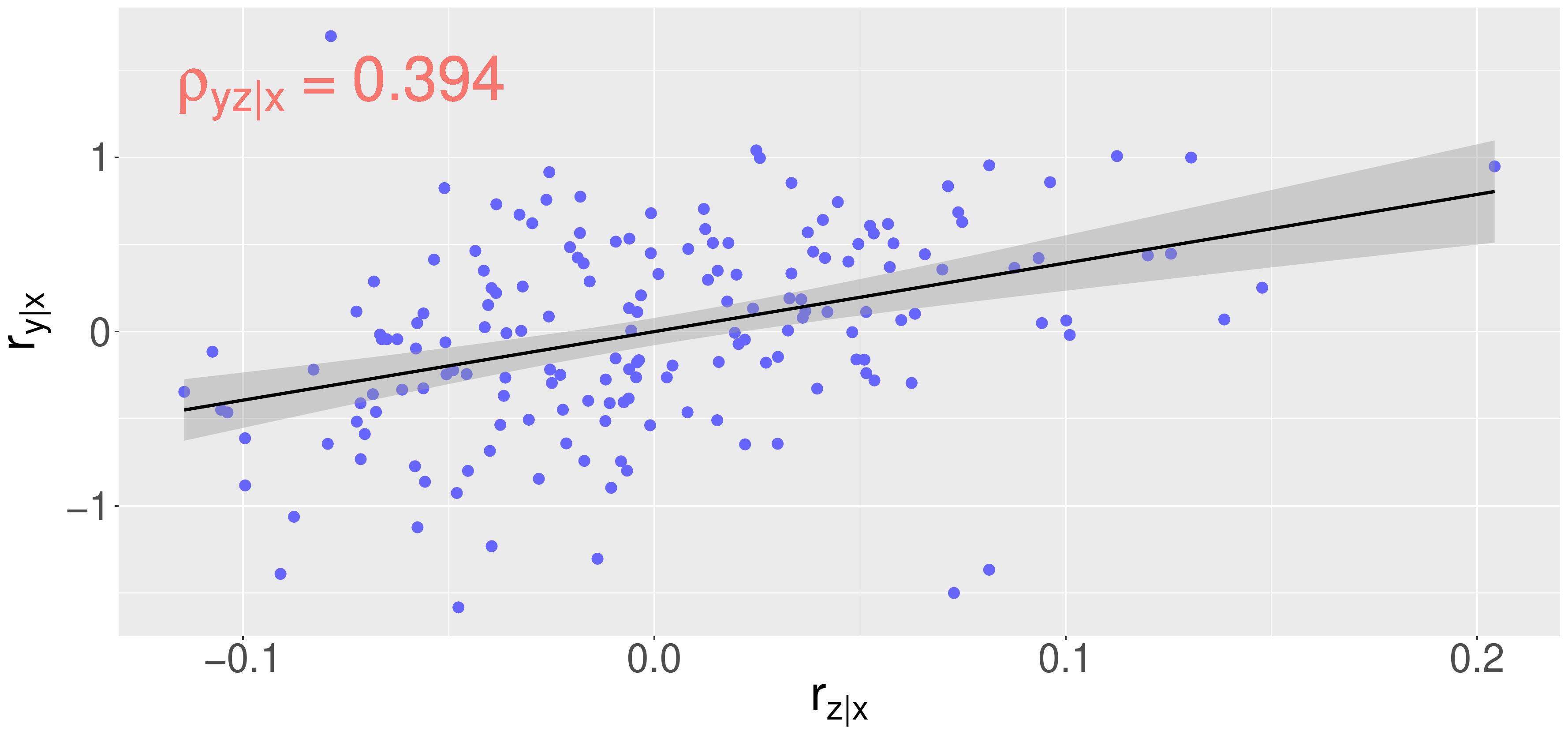}}\qquad\qquad
\subfigure[]
{\includegraphics[scale=0.22]{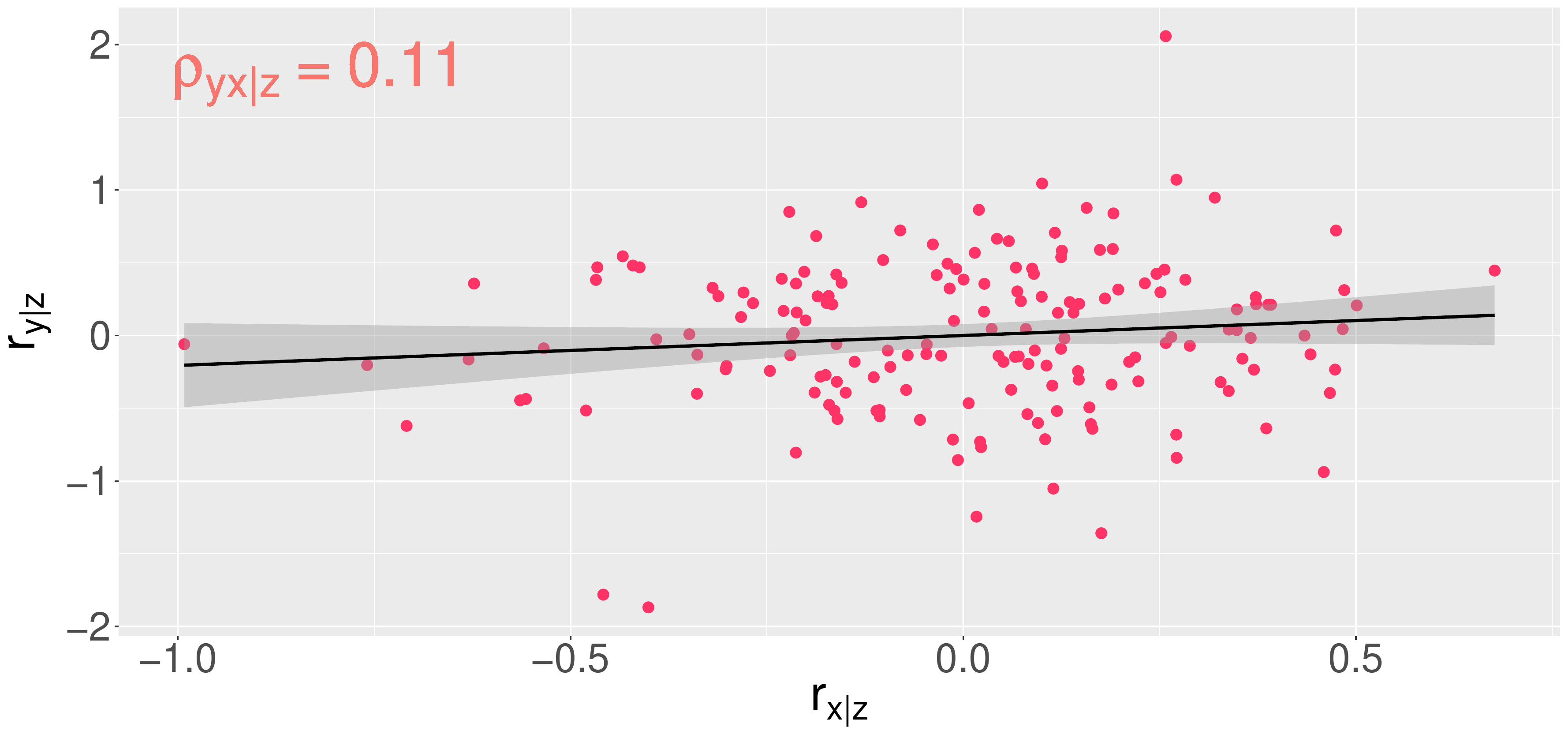}}\qquad\qquad
\subfigure[]
{\includegraphics[scale=0.22]{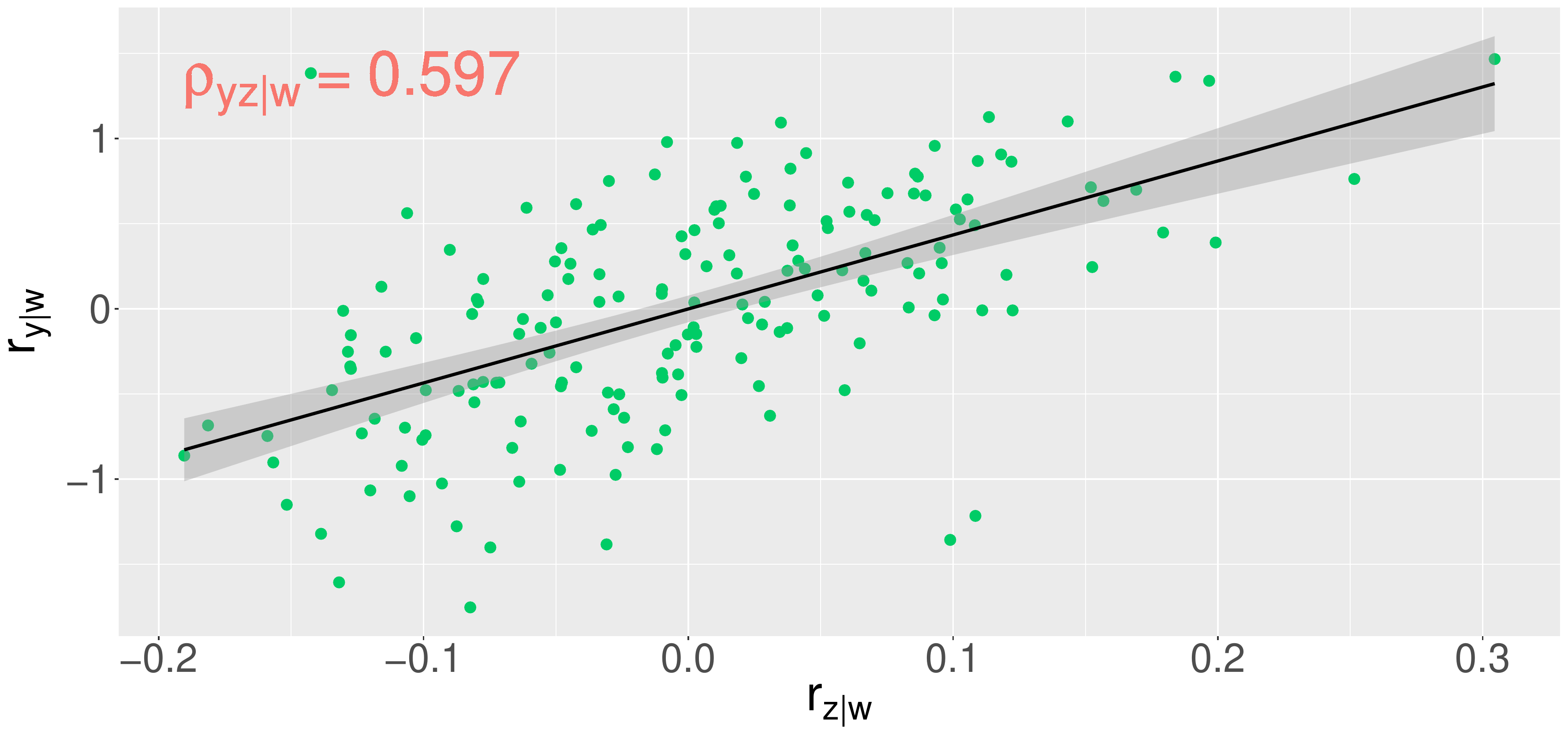}}
\caption{Correlation between residuals for the 1st van de Bosch's Sample, with $y=M_{\bullet}$,  $x=M_{G}\sigma^2$, $z=\sigma$, $w=M_G$: (a) $\rho_{yz|x}$, (b) $\rho_{yx|z}$ and (c) $\rho_{yz|w}$. }
\label{fig:vdb174}
\end{figure*}

\begin{figure*}[!h]
\centering%
\subfigure[]
{\includegraphics[scale=0.22]{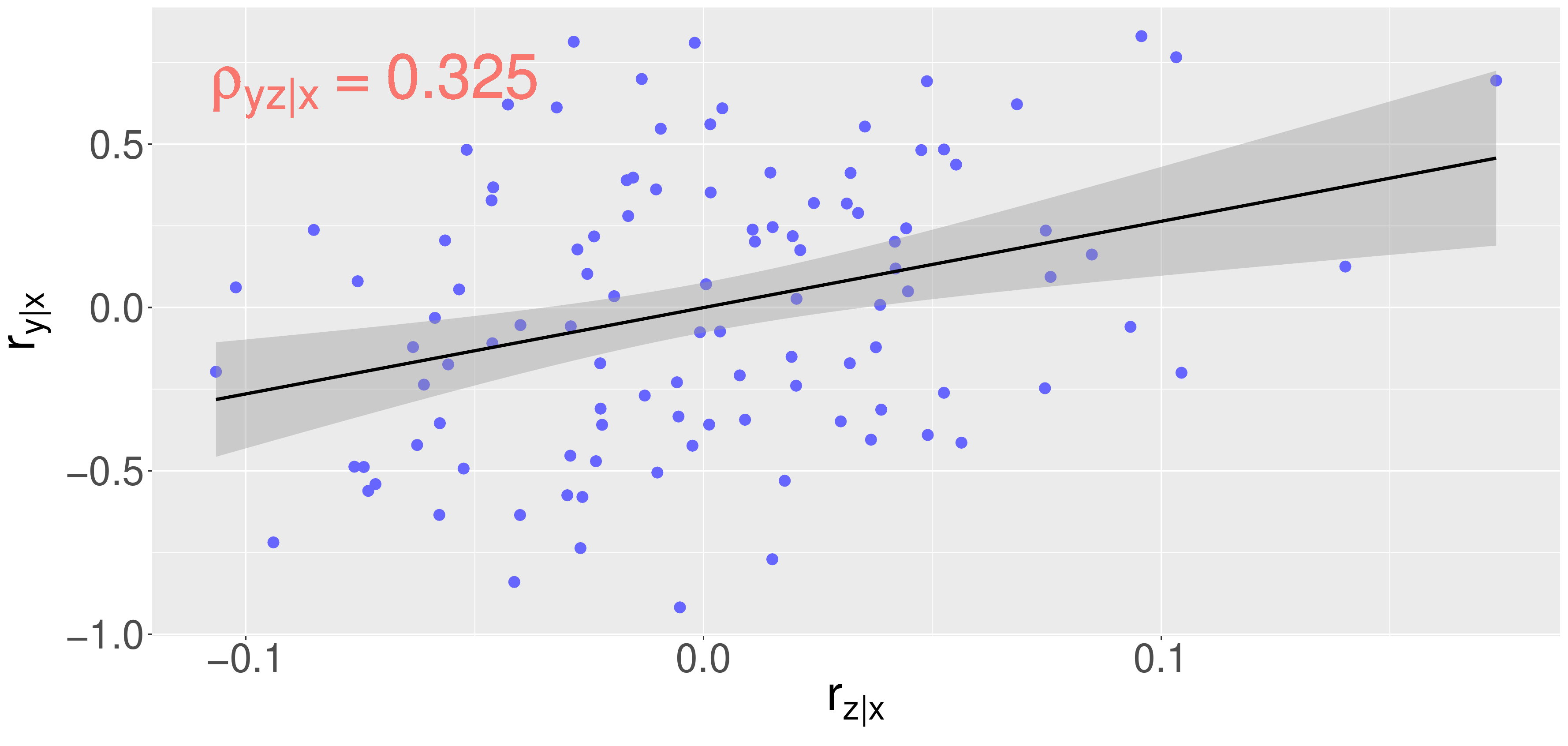}}\qquad\qquad
\subfigure[]
{\includegraphics[scale=0.22]{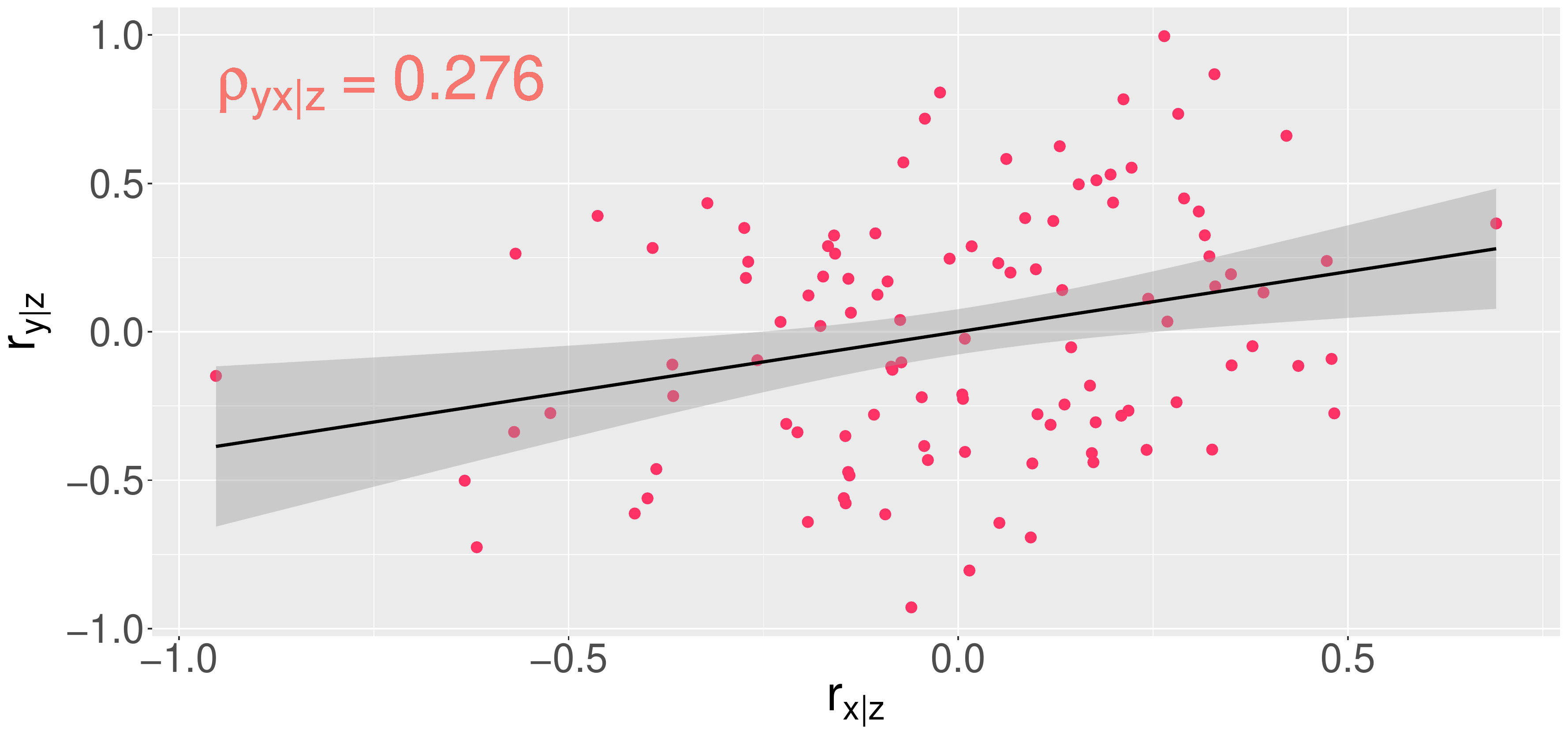}}\qquad\qquad
\subfigure[]
{\includegraphics[scale=0.22]{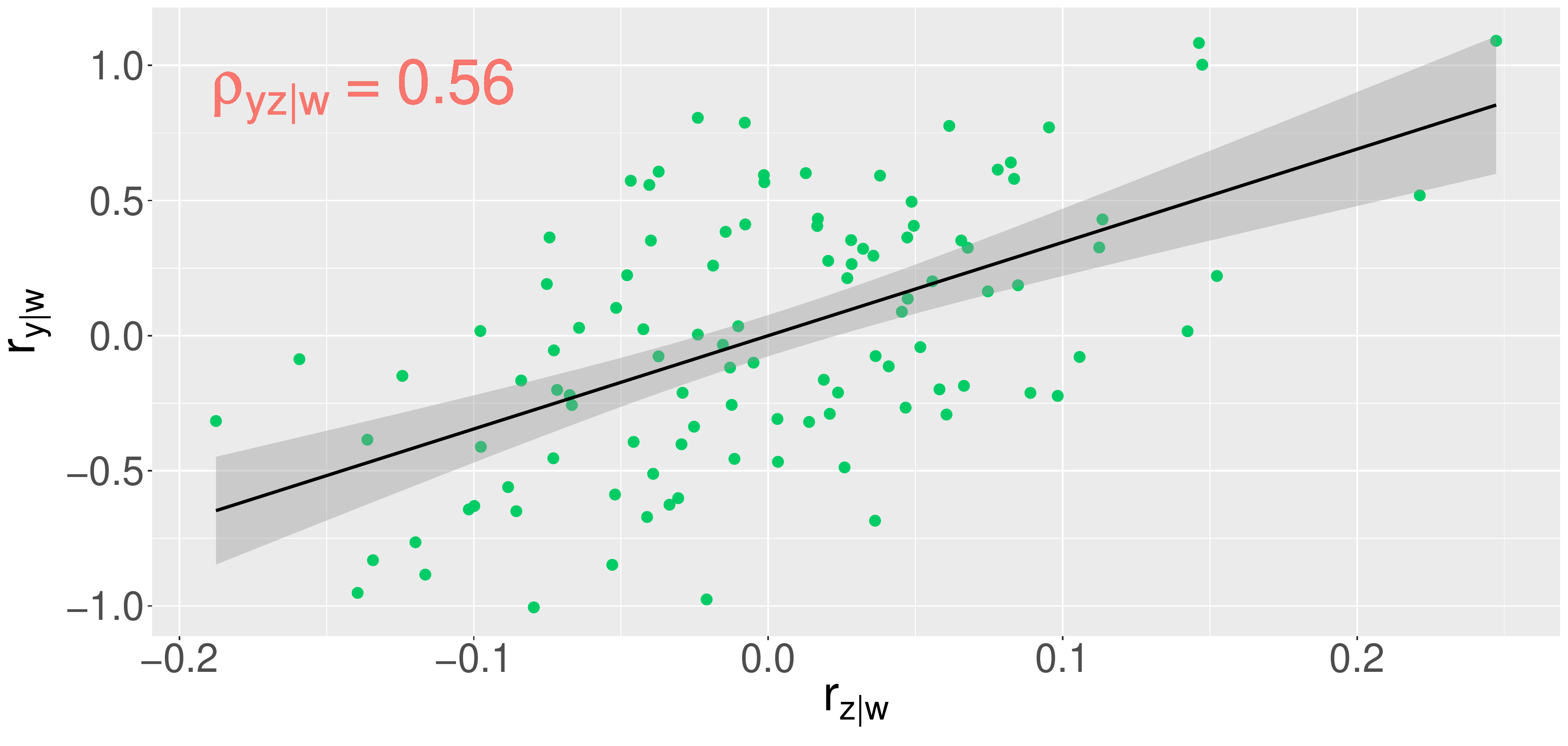}}
\caption{Correlation between residuals for the 2nd van de Bosch's Sample, with $y=M_{\bullet}$,  $x=M_{G}\sigma^2$, $z=\sigma$, $w=M_G$: (a) $\rho_{yz|x}$, (b) $\rho_{yx|z}$ and (c) $\rho_{yz|w}$.}
\label{fig:vdb108}
\end{figure*}

\begin{figure*}[!h]
\centering%
\subfigure[]
{\includegraphics[scale=0.22]{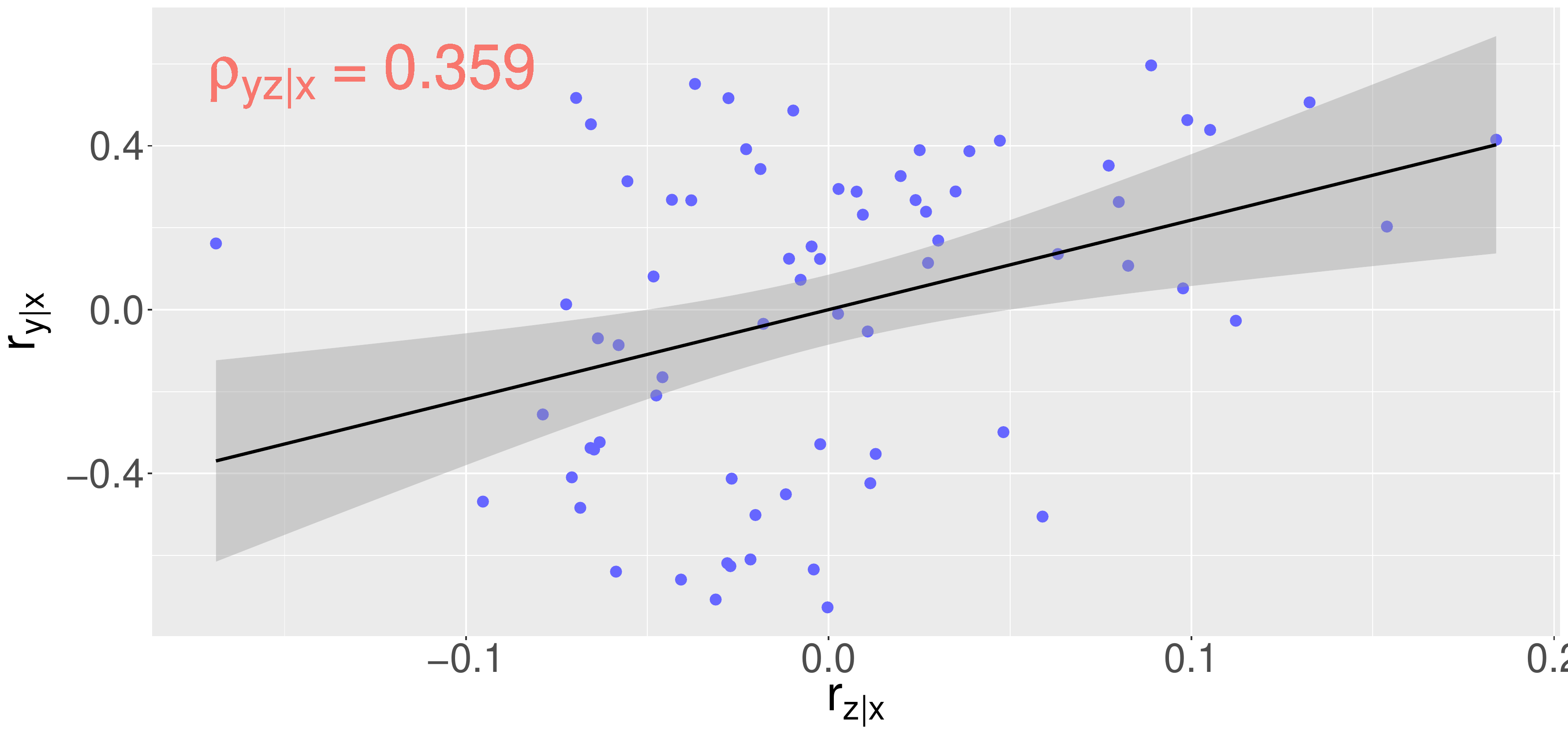}}\qquad\qquad
\subfigure[]
{\includegraphics[scale=0.22]{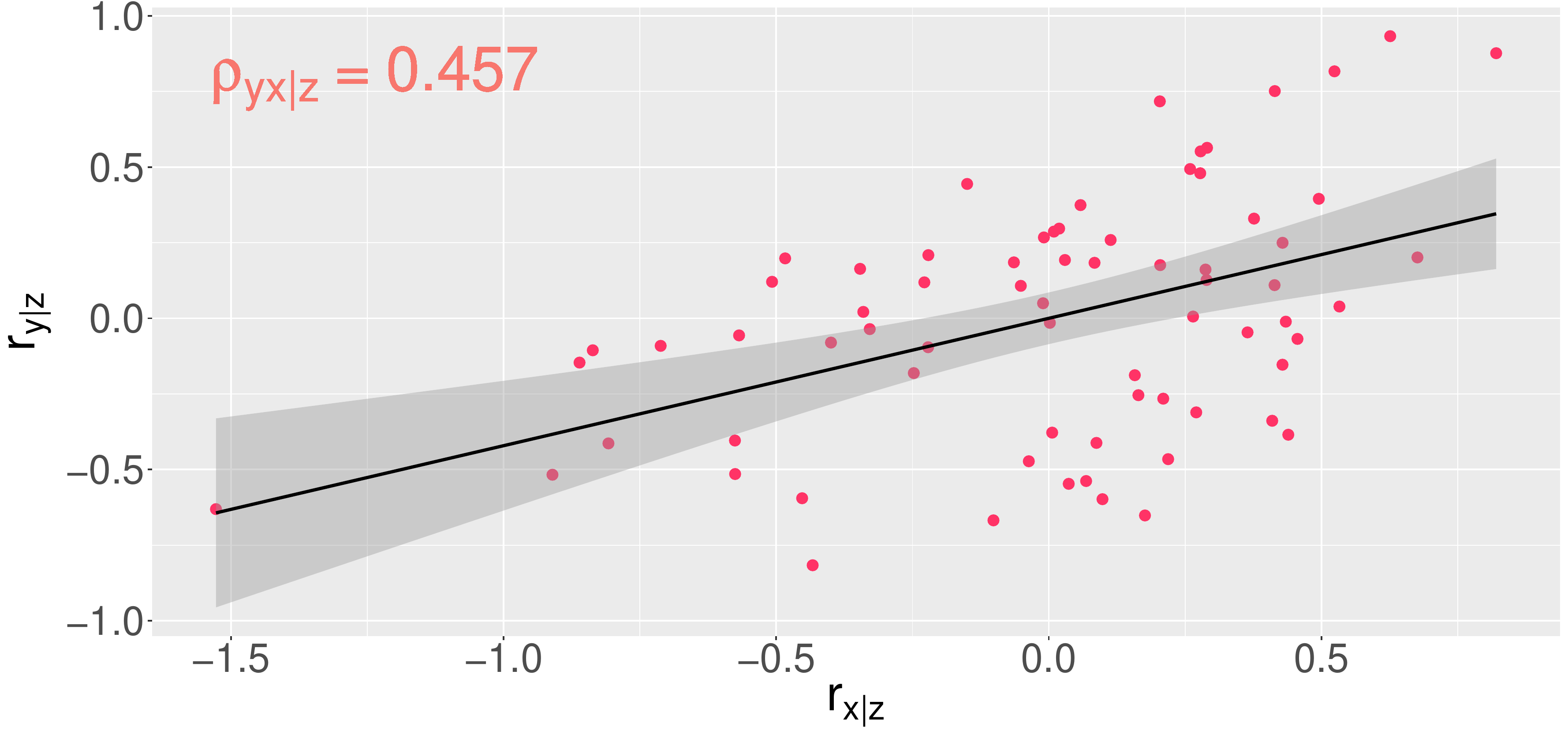}}\qquad\qquad
\subfigure[]
{\includegraphics[scale=0.22]{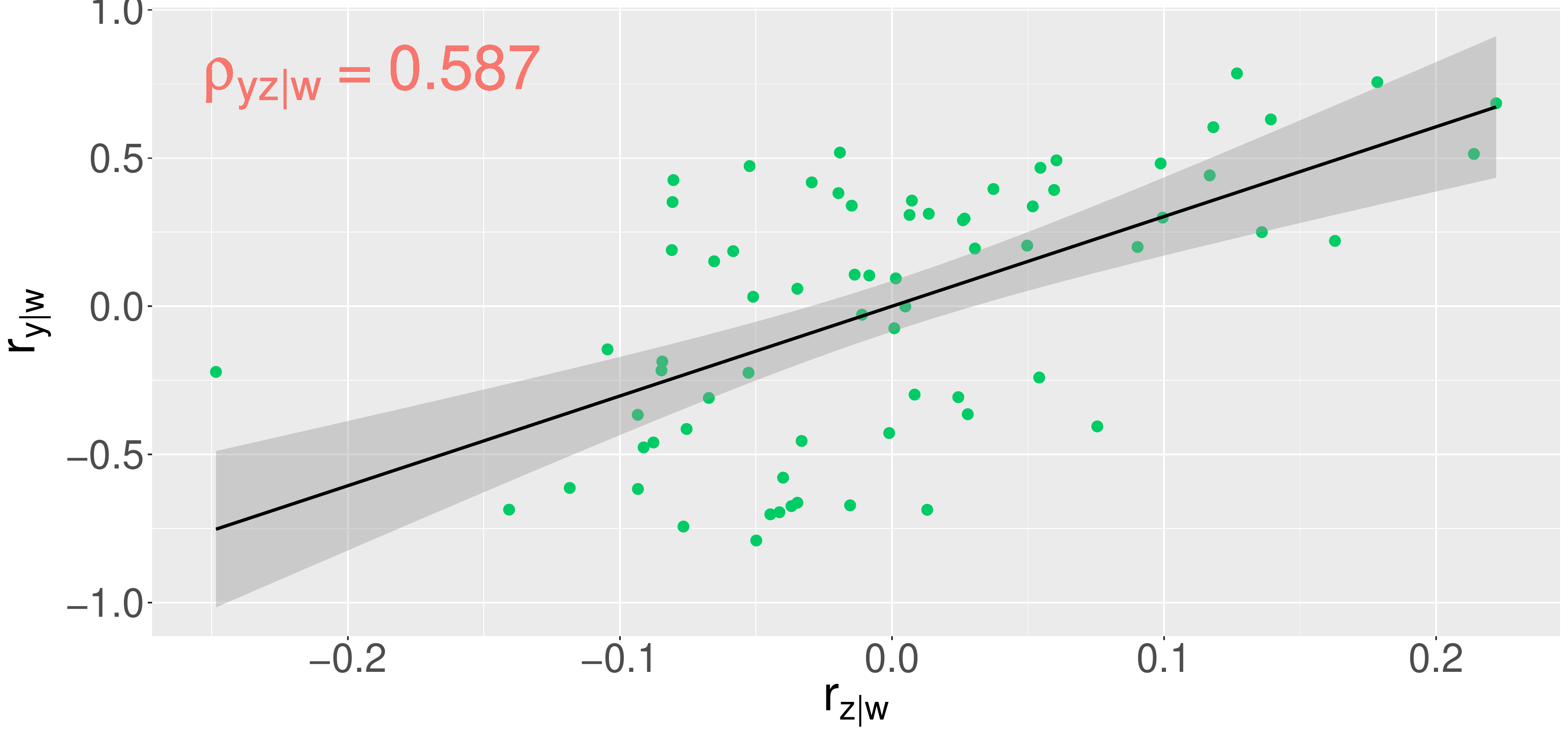}}
\caption{Correlation between residuals for the de Nicola- Saglia' Sample, with $y=M_{\bullet}$,  $x=M_{G}\sigma^2$, $z=\sigma$, $w=M_G$: (a) $\rho_{yz|x}$, (b) $\rho_{yx|z}$ and (c) $\rho_{yz|w}$.}
\label{fig:ns}
\end{figure*}

\begin{figure*}[!h]
\centering%
\subfigure[]
{\includegraphics[scale=0.22]{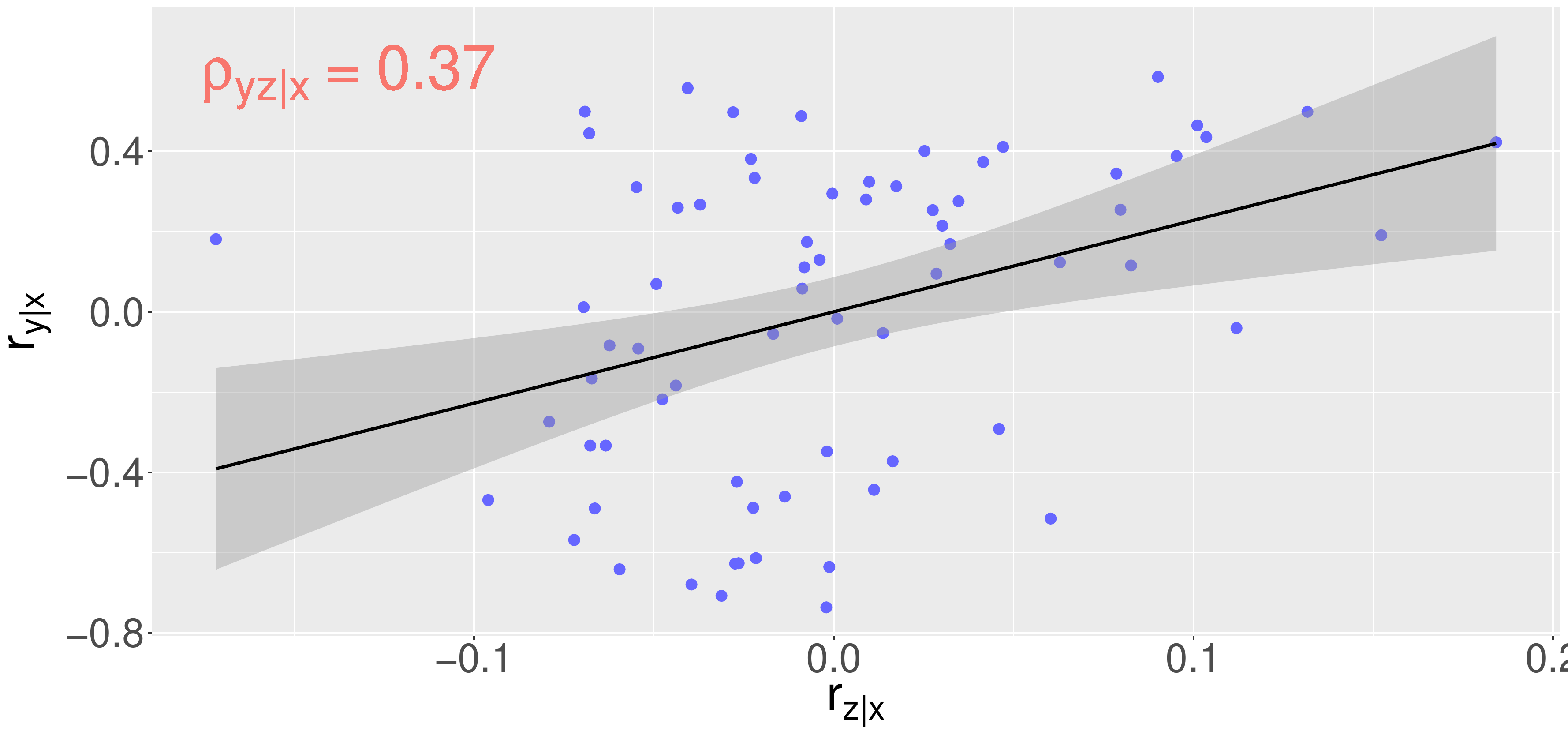}}\qquad\qquad
\subfigure[]
{\includegraphics[scale=0.22]{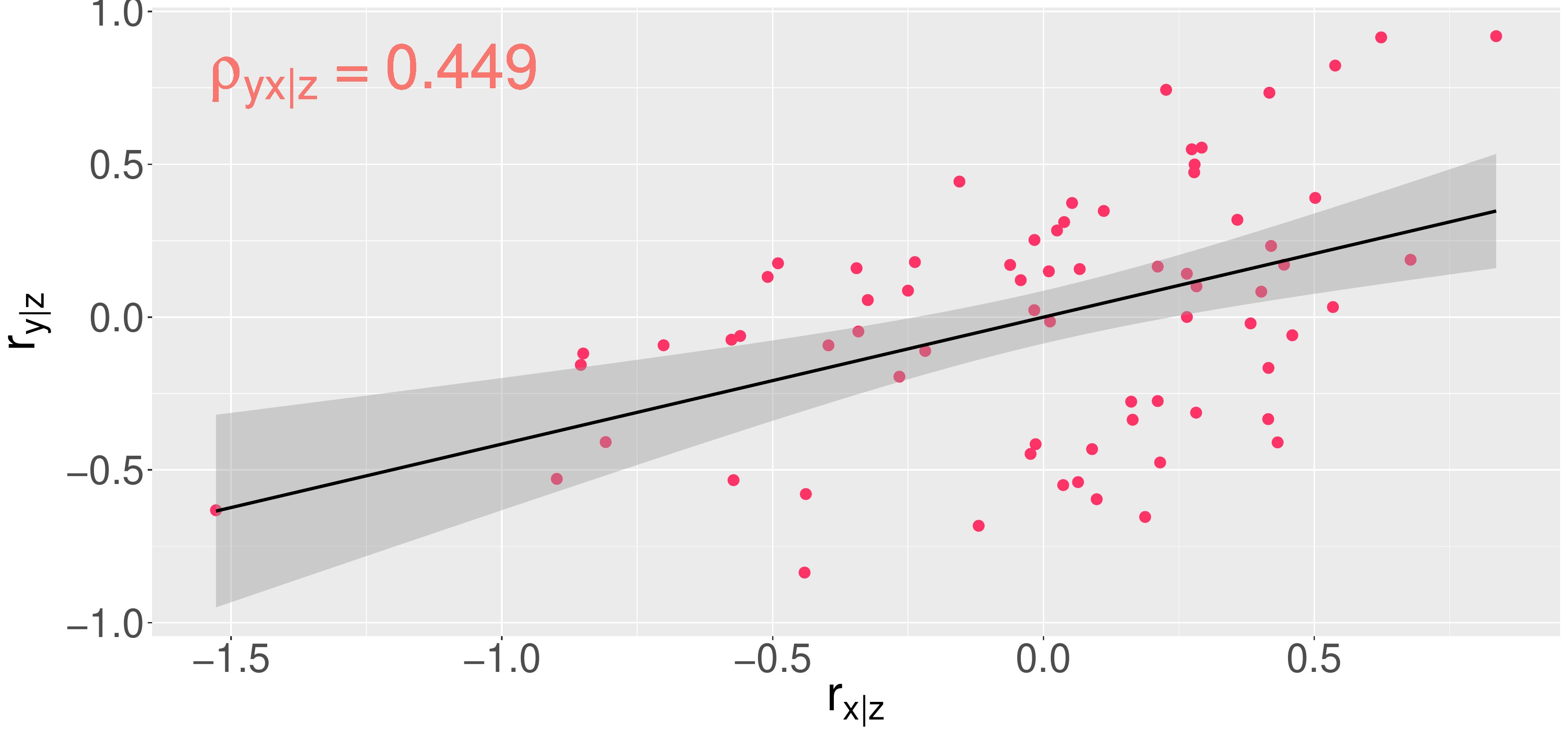}}\qquad\qquad
\subfigure[]
{\includegraphics[scale=0.22]{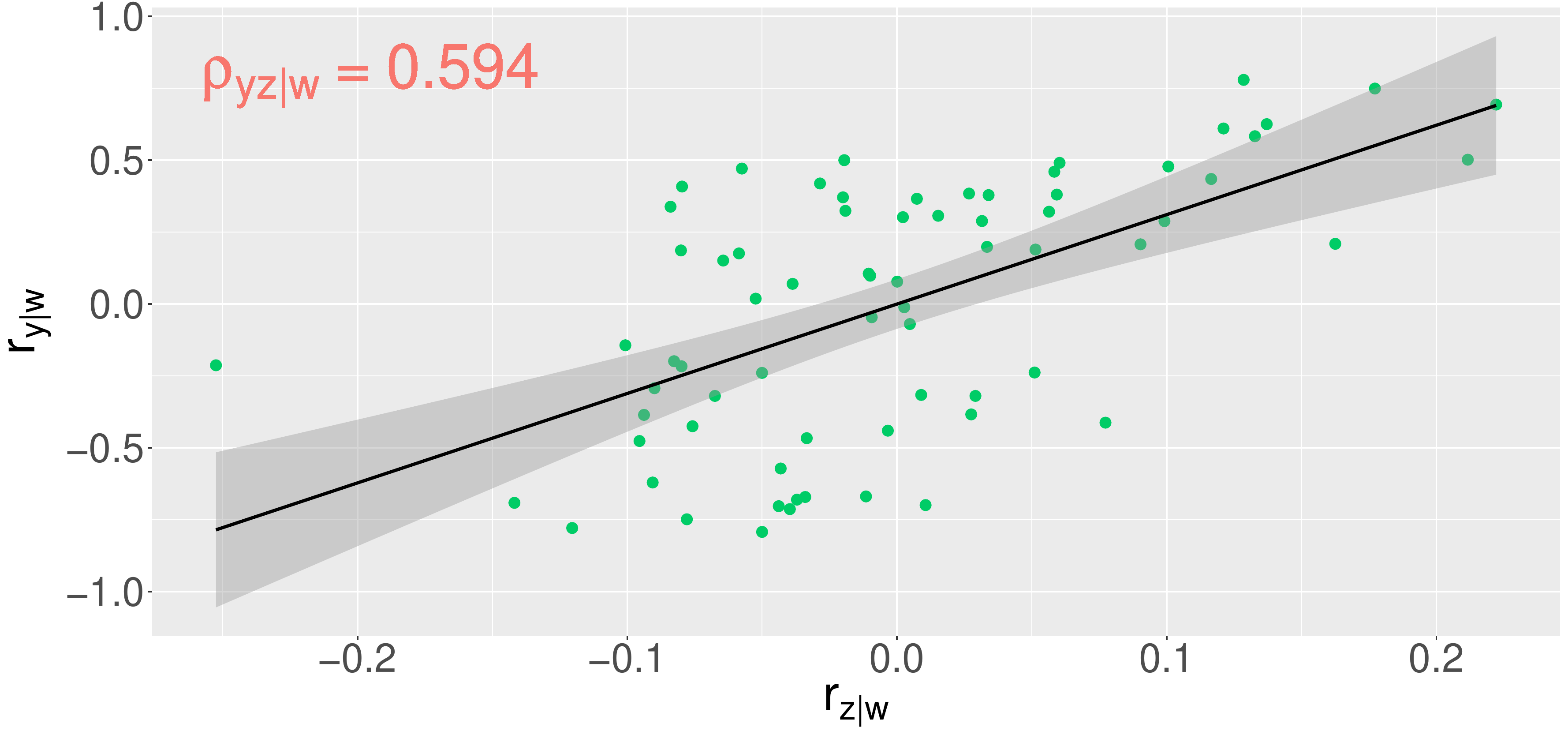}}
\caption{Correlation between residuals for the Saglia' Sample, with $y=M_{\bullet}$,  $x=M_{G}\sigma^2$, $z=\sigma$, $w=M_G$: (a) $\rho_{yz|x}$, (b) $\rho_{yx|z}$ and (c) $\rho_{yz|w}$.}
 \label{fig:sa}
\end{figure*}

Figures from \ref{fig:cap} to \ref{fig:kh} allow a visual inspection of conditional correlations among residuals from scaling relations. The fitted lines pass through the origin $(0,0)$ since the sets of residuals have null mean. The steeper the line the larger the linear correlation. The shaded area can be interpreted as a tolerance region that gives $95\%$ confidence intervals around the fitted line in a pointwise fashion.
In general, based on the fitted model (\ref{mod1F}), confidence intervals around the fitted line are obtained as $\hat\mu_i \pm t_{0.975; n-2}se(\hat\mu_i)$, $i=1,2,\ldots,n$, where  $t_{0.975; n-2}$ is the $0.975-$level quantile of the $t_{n-2}$ distribution and $se(\hat\mu_i)$ is the standard error associated to the fitted value $\hat\mu_i$. Let $X=[\mathbf{1} | x]$ be the $n \times 2$ design matrix, with $x=(x_1,x_2,\ldots,x_n)^\top$. The fitted variance-covariance matrix of $(\hat b, \hat m)$ is $\hat\epsilon_0^2 (XX^\top)^{-1}$. Then, $se(\hat\mu_i)=\hat\epsilon_0\sqrt{\mathbf{x}_i^\top (XX^\top)^{-1}\mathbf{x}_i}$, with $\mathbf{x}_i=(1,x_i)^\top$.
When the shaded area covers entirely the abscissa axis, as in the right panels of Fig. \ref{fig:vdb174}, it means there is not any evidence of correlation.

 \begin{table}[h]
	\caption{Conditional correlations with $95\%$ confidence intervals, with $y=M_{\bullet}$,  $x=M_G \sigma^2/c^2$, $w=M_G$ and $z=\sigma$, where $\rho_{yw|z}=\rho_{yx|z}$}
	\begin{center}
		\smallskip
		\begin{tabular}{r|ccc}
			\tableline
			\tableline
			Sample&\multicolumn{2}{c}{Cond. correl}&$95\%$ CI\\
			\tableline
			Cappellari&$\rho_{yz|w}$&0.604&0.384 $\div$ 0.760\\
			&$\rho_{yz|x}$&0.338&0.056 $\div$ 0.570\\
			&$\rho_{yx|z}$&0.341&0.059 $\div$ 0.572\\
			\tableline

			van den Bosch\_174	&$\rho_{yz|w}$&0.597&0.492 $\div$ 0.685\\
            &$\rho_{yz|x}$&0.394&0.260 $\div$ 0.512\\
			&$\rho_{yx|z}$&0.110&-0.040 $\div$ 0.254\\
			\tableline

			van den Bosch\_108	&	$\rho_{yz|w}$&0.560&0.415 $\div$ 0.678\\
            &	$\rho_{yz|x}$&0.325&0.145 $\div$ 0.484\\
		    &	$\rho_{yx|z}$&0.276& 0.092 $\div$ 0.442\\
			\tableline

			De Nicola-Saglia &$\rho_{yz|w}$&0.587&0.409 $\div$ 0.721\\
            &$\rho_{yz|x}$&0.359&0.138 $\div$ 0.547\\
			&$\rho_{yx|z}$&0.457& 0.251 $\div$ 0.624\\
			\tableline
			
			Saglia 	&$\rho_{yz|w}$&0.593&0.419 $\div$ 0.727\\
            &$\rho_{yz|x}$&0.370&0.150 $\div$ 0.555\\
			&$\rho_{yx|z}$&0.449&0.241 $\div$ 0.618\\
			\tableline

            Kormendy-Ho	&$\rho_{yz|w}$&0.566&0.327 $\div$ 0.737\\
            &	$\rho_{yz|x}$&0.208&-0.092 $\div$ 0.472\\
			&	$\rho_{yx|z}$&0.576&0.340 $\div$ 0.744\\
            \tableline
		\end{tabular}
	\end{center}
	\label{cor7}
\end{table}

 The detailed results for each sample are given below.
\begin{itemize}
	\item[$\bullet$] \textit{Cappellari}. The entries in Table \ref{cor7} and Fig. \ref{fig:cap} show that there is no evidence of any difference between the magnitude of the two correlations coefficients. According to Table \ref{F6}, we have that both variables provide a significant reduction at a significance level not lower than about $2\%$.
	In summary, in this case, both $M_{JAM}\sigma^2$ and $\sigma$ share a close relative importance.
	
     \vspace{5pt}
	
	\item[$\bullet$] \textit{van den Bosch\_174}. From the inspection of conditional correlations in Table \ref{cor7} and Fig. \ref{fig:vdb174}, we have that kinetic energy is not significantly correlated with black hole mass given dispersion velocity, whereas there is evidence of a moderate conditional correlation between black hole mass and dispersion velocity given kinetic energy. The analysis in Table \ref{F6} confirms that kinetic energy is less important than the dispersion of velocity, whereas, in contrast, the inclusion of velocity dispersion improves significantly over the $M_{\bullet}-M_{dyn}\sigma^2$ relation.
	
     \vspace{5pt}
	
	\item[$\bullet$] \textit{van den Bosch\_108}.  From the inspection of conditional correlations in Table \ref{cor7} and Fig. \ref{fig:vdb108}, we have that the relative importance of kinetic energy is lower than velocity dispersion, whereas there isn't evidence of a difference in conditional correlations. The analysis in Table \ref{F6} confirms that kinetic energy has a slightly lower importance, since it determines a smaller, but significant still, reduction in the sum of squares. Here, the two explanatory variables under study can be claimed to be equivalent even if $\sigma$ gives a slightly larger contribute.
	
     \vspace{5pt}
	
	\item[$\bullet$] \textit{de Nicola - Saglia}.  From the inspection of conditional correlations in Table \ref{cor7} and Fig. \ref{fig:ns}, we have that the relative importance of kinetic energy is larger than velocity dispersion, whereas there isn't evidence of a difference in conditional correlations.  The analysis in Table \ref{F6} confirms that kinetic energy has a slightly larger importance, since it determines a larger significant reduction in the sum of squares.
	
     \vspace{5pt}
	
	\item[$\bullet$] \textit{Saglia}.  From the inspection of conditional correlations in Table \ref{cor7} and Fig. \ref{fig:sa}, we have,  the velocity dispersion exhibits a larger conditional correlation with the mass of SMBH, but there is no evidence of a significant difference with the conditional between black hole mass and kinetic energy at a $5\%$ significance level. The reduction in the residual sum of squares is given in Table \ref{F6}. The introduction of kinetic energy improves the $M_{\bullet}-\sigma$ relation with a larger reduction.

     \vspace{5pt}
	
	\item[$\bullet$] \textit{6th Sample: Kormendy-Ho}.  From the inspection of conditional correlations in Table \ref{cor7} and Fig. \ref{fig:kh}, we can note that there is evidence of a larger conditional correlation between black hole mass and kinetic energy given dispersion velocity, while velocity dispersion is not significantly correlated with black hole mass given the kinetic energy. From the values that we can observe in Table \ref{F6}, it is possible to note that kinetic energy is more important than the velocity dispersion and improves significantly over the $M_{\bullet}-\sigma$ relation.

\end{itemize}

\begin{table}[!h]
	\caption{Residual sum of squares reductions $\Delta$ and corresponding p-values.}
	\begin{center}
		\smallskip
		\begin{tabular}{@{}r|ccc@{}}
			\tableline
			\tableline
			Sample&& $\Delta$ & p-value \\
			\tableline
		Cappellari	&$\displaystyle M_{JAM}\sigma^2 \over \displaystyle c^2$ & 0.893 &  0.0205 \\
		&	$\sigma$ & 0.877& 0.0216 \\
			\tableline
		van den Bosch\_174	&	$\displaystyle M_{dyn}\sigma^2 \over \displaystyle c^2$ & 0.573 &  0.151 \\
		&	$\sigma$ & 8.629& $<$0.001 \\
			\tableline
		van den Bosch\_108&	$\displaystyle M_{dyn}\sigma^2 \over \displaystyle c^2$ & 1.399 &   0.004 \\
		&	$\sigma$ & 2.002&  $<$0.001 \\
			\tableline
		De Nicola-Saglia&	$\displaystyle M_{Bu}\sigma^2 \over \displaystyle c^2$ & 2.373  &  $<$0.001 \\
		&	$\sigma$ & 1.331& 0.002 \\
			\tableline
		Saglia&	$\displaystyle M_{Bu}\sigma^2 \over \displaystyle c^2$& 2.315 &  $<$0.001 \\
		&	$\sigma$ & 1.452& 0.002 \\
			\tableline
		Kormendy- Ho&	$\displaystyle M_{Bu}\sigma^2 \over \displaystyle c^2$& 1.532 &  $<$0.001 \\
		&	$\sigma$ & 0.139& 0.177 \\
			\tableline
		\end{tabular}
	\end{center}
	\label{F6}
\end{table}

\section{Conclusions}
In this paper we shed more light on the role played by the single variables in the study of scaling relations and on the correct interpretation of some common statistical analyses. In addition, we investigated the considered scaling relations not only from a goodness of fit point of view but also under a predictive perspective.
We have compared the performance of two well known relations  the $M_{\bullet} - \sigma$ and the $M_{\bullet} - M_{G}\sigma^2$. It has been shown that both relations work satisfactory with all the six samples analyzed in this paper. In particular with the samples one, three, four and five the behavior of the two relations can be considered almost equivalent, while with the sample two the first law leads to better results and with sample six the opposite occurs.  The comparison from a predictive point of view gives more relevance to the reliability of the $M_{\bullet} - M_{G}\sigma^2$ relation and the residual analysis did not unveil any substantial preference of one variable, the kinetic energy, over another, the velocity dispersion. It is worth noting that simple scaling relations may be significantly improved by considering a suitably chosen extra variable, for instance introducing the kinetic energy in the  $M_{\bullet} - \sigma$, while the same does not occur considering the mass of the galaxy as the extra parameter, in agreement with previous work.

\section*{\footnotesize Acknowledgements}
{\footnotesize This research was partially supported by FAR fund of the University of Sannio.}

\end{document}